\documentclass[aps,pre,onecolumn,longbibliography,nofootinbib,showkeys,showpacs,amssymb,floatfix]{revtex4-1}

\usepackage{latexsym}
\usepackage{amsfonts}
\usepackage{color}
\usepackage{graphicx}
\usepackage{mathptmx}      
\usepackage{bm}
\usepackage{enumerate}
\usepackage{subfigure}
\usepackage{listings}
\usepackage{grffile}

\definecolor{darkgreen}{rgb}{0.00,0.39,0.00}

\newcommand\sign{\mathop{\mathrm{sign}}}
\newcommand\hidden{\phi}

\begin{document}




\title{The photon identification loophole in EPRB experiments: \\computer models with single-wing selection}

\author{H. De Raedt}
\email{h.a.de.raedt@rug.nl}
\thanks{Corresponding author}
\affiliation{Zernike Institute for Advanced Materials,\\
University of Groningen, Nijenborgh 4, NL-9747AG Groningen, The Netherlands}
\author{K. Michielsen}
\affiliation{Institute for Advanced Simulation, J\"ulich Supercomputing Centre,\\
Forschungszentrum J\"ulich, D-52425 J\"ulich, Germany}
\affiliation{RWTH Aachen University, D-52056 Aachen, Germany}

\author{Karl Hess}
\affiliation{Center for Advanced Study, University of Illinois, Urbana, Illinois, USA}

\date{\today}

%
%

\begin{abstract}
Recent Einstein-Podolsky-Rosen-Bohm experiments [M. Giustina {\it et al}. Phys. Rev. Lett. 115, 250401 (2015);
L. K. Shalm {\it et al}. Phys. Rev. Lett. 115, 250402 (2015)]
that claim to be loophole free are scrutinized and are shown to suffer a photon identification loophole.
The combination of a digital computer and discrete-event simulation
is used to construct a minimal but faithful model of the most perfected realization of these laboratory experiments.
In contrast to prior simulations, all photon selections are strictly made, as they are in the actual experiments,
at the local station and no other ``post-selection'' is involved.
The simulation results demonstrate that a manifestly non-quantum model that identifies photons
in the same local manner as in these experiments can produce correlations that
are in excellent agreement with those of the quantum theoretical description
of the corresponding thought experiment, in conflict with Bell's theorem.
The failure of Bell's theorem is possible because of our recognition of the photon identification loophole.
Such identification measurement-procedures are necessarily included in all actual experiments
but are not included in the theory of Bell and his followers.
\end{abstract}



\maketitle


%
\begin{figure}[ht]
\begin{center}
\includegraphics[width=0.95\textwidth]{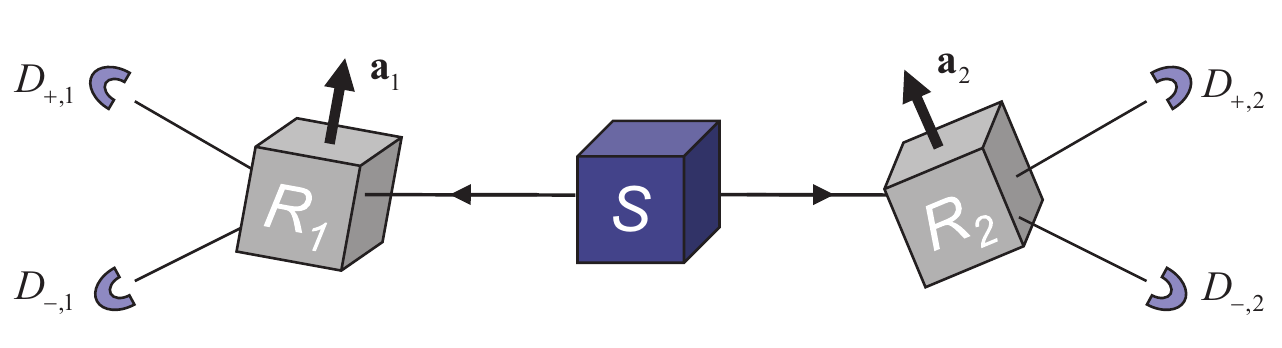}
\caption{%
Diagram of the EPRB thought experiment.
The source $S$, activated at times labeled by $n=1,2,\ldots,N$, sends a photon to the observation station
$R_1$ and another photon to the observation station $R_2$.
Depending on the settings of these observation stations
represented by unit vectors $\mathbf{a_1}$ and $\mathbf{a_2}$,
the signal going to the left (right) triggers the counters $D_{+,1}$ or $D_{-,1}$ ($D_{+,2}$ or $D_{-,2}$).
}
\label{fig0}
\end{center}
\end{figure}

The debate between Einstein and Bohr, about the foundations of quantum mechanics, resulted in a Gedanken- experiment suggested
by Einstein-Podolsky-Rosen (EPR)~\cite{EPR35} that was later modified by Bohm (EPRB)~\cite{BOHM51}.
The schematics of this experiment is shown in Fig.~\ref{fig0}
and involves two wings and two measurement stations. EPR used the quantum mechanical predictions for possible outcomes of
this experiment to show that quantum mechanics was incomplete.

Many years later, John S. Bell derived an inequality for the possible outcomes of EPRB experiments that he perceived to be based
only on the physics of Einstein's relativity~\cite{BELL01}. Bell's inequality, as it was henceforth called, seemed to contradict the quantum
results for EPRB experimental outcomes altogether~\cite{BELL01}.

Experimental investigations, following Bell's theoretical suggestions, provided a large number of data that were violating
Bell-type inequalities and climaxed in the suspicion of a failure of Einstein's physics and his basic understanding of space and
time~\cite{KOCH67,CLAU78,ASPE82b,WEIH98,CHRI13}.

Central to these discussions and questions are the correlations of space-like separated detection events,
some of which are interpreted as the observation of a pair of entities such as photons.
The problem of classifying events as the observation of a ``photon'' or of something else
is not as simple as in the case of say, billiard balls.
The particle identification problem is, in fact, key for the understanding of the epistemology of correlations between events.

What do we know about such correlations of space-like separated events?
Popular presentations of Bell's work typically involve two isolated persons (Alice, Bob) at separated
measurement stations (Tenerife, La Palma), who just collect data of local measurements.
But how does Alice know that she is dealing with a particle of a pair of which Bob investigates the other particle?
She is supposed to be totally isolated from Bob's wing of the experiment in order to fulfill Einstein's separation and locality principle!
The answer is that neither Alice nor Bob know they deal with correlated pairs if their stations are completely separated from each other
and have no space-time knowledge of the other wing ever.

In his theoretical work on the EPRB experiment, Bell did not address this fundamental question
but considered correlated pairs as given, without any trace of the tools of measurement and of space-time concepts
that are both necessary to accomplish the identification of events.
He then claimed to have discovered a conflict between his theoretical description
and the quantum theoretical description of the EPRB thought experiment~\cite{BELL01}.
As a consequence of this discovery, much research was devoted to
\begin{enumerate}[(i)]
\item
the actual derivation of Bell-type inequalities
from Einstein's framework of physics (particularly his separation principle that derives from the speed of light in vacuum
($c$ being the limit of all speeds) and Kolmogorov's probability theory~\cite{KOLM56},
\item
designing and performing laboratory experiments that provide data that are in conflict with the Bell-type inequalities,
\item
constructing mathematical-physical models, at times supported by computer simulations that entirely comply with Einstein's
relativity framework and separation principle, that do not rely on concepts of quantum theory,
and are nevertheless in conflict with the Bell's theorem.
\end{enumerate}

Bell's theory, and the theories of all his followers, including Wigner, do not deal with the identification of the correlated
particles and assume that the measurement pairs corresponding to the correlated particles are known automatically, so to speak per fiat.
But the knowledge of pairing requires additional data or additional channels of information.
These additional data may be measurement times, certain thresholds for detection and many other elements of the physical reality
of the experiments in both wings.
It is important to note that these data must involve measurements in both stations and are necessarily influencing the possible
knowledge of the correlations of the single measurements in these stations.  In the case of atomic or subatomic measurements the
measurement equipment does not only influence the single outcomes as Bohr has taught us, but correlated measurement equipment
(such as synchronized clocks or instruments that determine thresholds) also influence the knowledge of
correlations of these single measurements.
It is this extension of the Copenhagen view that leads to a loophole in Bell's Theorem, the photon identification loophole.
The violations of Bell-type inequalities described in this paper are based on this loophole.

Bell and followers envisage that the correlations may be measured in the laboratory in complete separation and,
therefore, physical models of the Bell opponents must only use the measurements of two completely separated wings
operated by Alice and Bob who know nothing of each other.
As just explained, correlations of spatially separated events can only be conceived by involving the human-invented space-time
system in order to demonstrate the pairing, the knowledge that measurements of particles belong together.
Therefore, correlations can only be determined in a self-consistent way under the umbrella of a given space-time system that encompasses
the two or more measurement stations.
This space-time system that all experimenters subscribe to enables us to ban spooky influences out of science,
particularly for the EPR experiments that Einstein constructed for the purpose to show this fact.

We maintain that it is highly non-trivial to identify (correlated) photons by experimental methods
and that this identification involves, at least in some way, a space-time system, a system developed by
the human mind and agreed upon by all experimenters and evaluators of the Bell-type experiments.
In fact, the identification of particle pairs requires certain knowledge of the space-time properties of all the experimental equipment involved.
This knowledge must necessarily extend to measurement stations far from each other and is, therefore, ``non-local''.
Of course, this non-local knowledge does not imply that there are non-local physical influences on the data measured in the two wings.
In EPRB experiments~\cite{WEIH98,HENS15,GIUS15,SHAL15}, great care is taken to rule out the possibility that the
observed correlations are due to physical influences that travel with velocities not exceeding the speed of light in vacuum.
But without that non-local knowledge, only spooky influences are left as a possibility for connecting events in the two experimental wings.
A space-time knowledge of all involved equipment is nevertheless required to apply the scientific method.

\section{Aim and structure of the paper}\label{ORG}

As explained above, it is essential that the identification of photons is included into any meaningful theoretical model
of an EPRB experiment because otherwise, the model is too simple to describe this experiment.
Failing to do so, only spooky influences can explain the observed pair correlations.
Specifically, omitting the inclusion of data which select the photons and/or pairs opens an Einstein-local loophole,
which we call henceforth the photon identification loophole.
By design, Bell-type models for the recent experiments that claim to be loophole free~\cite{GIUS15,SHAL15}
suffer from this loophole.

The main aim of this paper is to show that by exploiting this loophole, or formulated more positively,
by constructing a model that captures the essence of these recent laboratory experiments~\cite{GIUS15,SHAL15},
a manifestly non-quantum computer simulation of an Einstein-local model
that employs the same local photon identification method in each wing of the EPRB experiments
as in recent EPRB experiments~\cite{GIUS15,SHAL15},
yields the photon pair correlation of a pair of photons in the singlet state (of their polarizations),
in blatant contradiction with Bell's theorem.

The paper is structured as follows.
Section~\ref{MET} argues that a simulation on a digital computer is a perfect laboratory experiment
with a physically existing device and can therefore be used as a metaphor for other laboratory experiments.
The material in this section forms the conceptual basis for developing computer simulation models of the
recent EPRB experiments~\cite{GIUS15,SHAL15}.

Section~\ref{COU} discusses the relevance of counterfactual definiteness (CFD) in relation to
the derivation of Bell-type inequalities and hence also for computer simulation models that yield data
that are in conflict with these inequalities.
In Section~\ref{COM}, we introduce a CFD-compliant computer simulation model of the
recent EPRB experiments~\cite{GIUS15,SHAL15}.
We discuss the correspondence of the essential elements of the latter with those of
the simulation model but refrain from plunging into the details of the algorithm itself.

Section~\ref{EQU} gives a simple, rigorous proof that operationally but not conceptually,
photon identification in each wing by a local voltage threshold~\cite{GIUS15},
photon identification in each wing by a local time window,
and photon identification by time-coincidence counting are all mathematically equivalent.
In Section~\ref{MOD}, we discuss the consequences of excluding from a model,
at least one feature that is essential for an experiment to yield useful data.
We argue that Bell-type models, which are believed
to be relevant for the description of recent EPRB experiments~\cite{GIUS15,SHAL15},
suffer from the photon identification loophole in a dramatic manner.

Section~\ref{CFD} is devoted to a simple proof that the simulation model
introduced in Section~\ref{COM} is CFD-compliant.
In Section~\ref{BEL}, we give a simple derivation of the Bell~\cite{BELL93},
Clauser-Horn-Shimony-Holt (CHSH)~\cite{CLAU69}, Eberhard~\cite{EBER93},
and Clauser-Horn (CH)~\cite{CLAU74} inequalities
for real data, not for the imagined data produced by probabilistic models (which are discussed in the Appendix)
and in Section~\ref{INE} we present
a more general inequality that accounts for the local photon identification procedure, employed
in the laboratory experiments.

In section~\ref{ALG}, we specify the computer simulation algorithm and the simulation procedure in full detail.
A representative collection of simulation results is presented in section~\ref{RES}.
The main conclusion from these simulations is that a non-quantum model
that employs the same photon identification method in each wing of the EPRB experiments
as the one used in recent EPRB experiments~\cite{GIUS15,SHAL15}, reproduces
the results of quantum theory of the EPRB thought experiment.

In section~\ref{POS}, we argue that all EPRB laboratory experiments with photons
can be viewed as a tool to characterize the response of the observation stations,
leading to the conclusion that this response, in particular the local photon identification rate,
depends on the settings of these stations, consistent with the assumptions made in constructing the simulation model
and debunking the hypothesis that the observed pair correlations can be explained by non-local influences only.
The paper ends with section~\ref{CON} which contains our conclusions.

\begin{figure}[t]
\begin{center}
\includegraphics[width=0.7\textwidth]{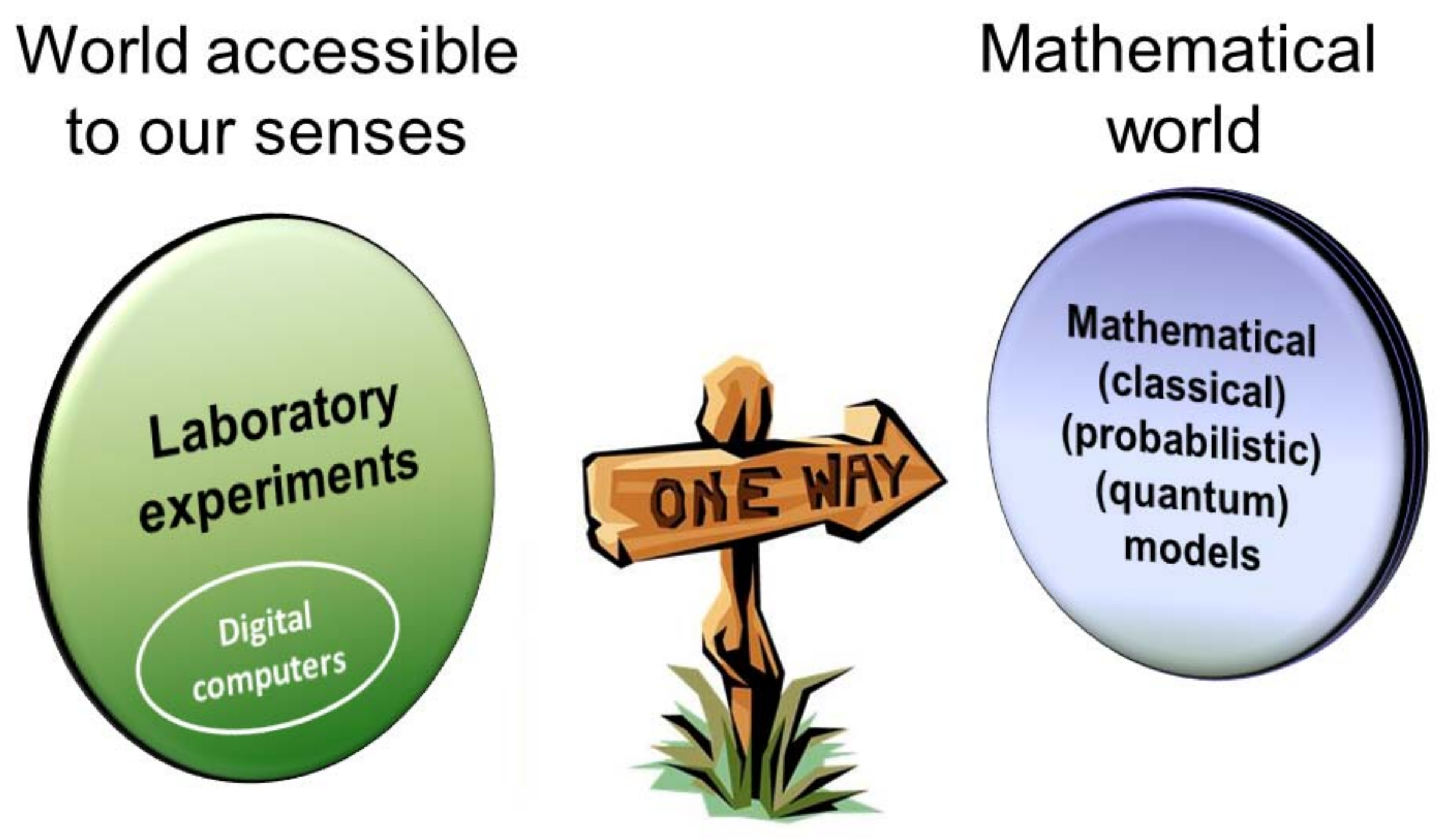}
\caption{%
Diagram indicating the direction of the modeling process adopted in this paper.
An algorithm executing on a digital computer is an instance of an experiment
on a physically existing device, a metaphor for a perfect laboratory experiment
in which there are no  unknown influences.
}
\label{fig0a}
\end{center}
\end{figure}

\section{Metaphor for a perfect laboratory experiment}\label{MET}

An important, characteristic feature of digital computers is that their logical operation does not depend on
the technology that is used to construct the machine.
These days, the first thing that comes to mind when talking about digital computers are the electronic machines based on semiconductor technology
but it is a fact that, although not cost-effective nor particularly useful in practice,
digital computers can also be built from mechanical parts, e.g. Lego\textsuperscript{TM} elements.

In former times, one could read off the state of the computer's internal registers from a LED display.
Although not practical at all, in principle one could use a huge LED display to show the internal state of the whole computer.
This is only to say that there is a one-to-one mapping from the state of the computer to sense impressions (e.g. light on/off).
Therefore, the metaphor also offers unique possibilities to confront man-made concepts
and theories with actual facts, i.e. real perfect experiments, because
it guarantees that we have a well-defined, precise
representation of the concepts and algorithms (both in terms of bits) involved
that directly translate into sense impressions.

In the analysis of laboratory EPRB experiments, it is essential
that all the important degrees of freedom that affect the data analysis are identified and included,
otherwise the conclusions drawn from an incomplete analysis may be wrong~\cite{GRAF15}.
Computer simulation puts us in the position to perform experiments under the same
mathematical conditions for which e.g. Bell-type inequalities can be derived,
simply because we can carry out real, perfect experiments that are void
of any unknown elements that may affect the results and analysis.

A digital computer is a physical (electronic or mechanical) device that changes its physical state (by flipping bits)
according to well-defined rules (the algorithm).
Therefore, assuming that the machine is operating flawless for the time period of interest
(a very reasonable assumption these days),
executing an algorithm on a digital computer is a physics experiment
in which there are no unknown elements of physical reality that might affect the outcome.
In this sense, the ``digital computer + algorithm'' can be viewed as a metaphor for a perfected laboratory experiment,
a discrete-event simulation that represents the so called ``loophole-free'' EPRB experiments~\cite{GIUS15,SHAL15}.

Starting with Bell's work~\cite{BELL93}, most theoretical work on the subject matter is based on probability theory.
This mathematical framework contains conceptual elements (probability measures and infinitesimals)
that are outside the domain of our sensory experiences and have no counterpart in our physical world.
Therefore, to avoid pitfalls, we first devise an algorithm
that simulates the perfect laboratory experiment and then construct a probabilistic model of this algorithm.
A graphical representation of the modeling philosophy that we adopt in this paper is shown in Fig.~\ref{fig0a}.
Note that our approach starts from the experiments and results in a theory, which we believe is the only direction one should go.
In contrast, Bell's approach was the design of an experiment starting from his theoretical point of view.

\section{Counterfactual definiteness}\label{COU}

Counterfactual reasoning~\cite{HESS16b} plays a significant role in the literature related to Bell's work and
is seen by many a conditia sine qua non to derive Bell-type inequalities. However, as explained below, the actual EPRB
experiments do not permit any proof of CFD compliance. This fact demonstrates an unexpected conceptual advantage of computer
experiments. We can turn on and off CFD compliancy at will in our algorithm and simulate the consequences and thus distinguish
the precise conditions that may or may not lead to violations of Bell-type inequalities. This is our reason to dedicate
significant sections of this paper to counterfactual definiteness (CFD) as defined below, and include CFD-compliant models at
the side of more faithful models of the actual EPRB experiments.

So called counterfactual ``measurements'' yield values that have been derived by means other
than direct observation or actual measurement, such as by calculation on the basis of a well-substantiated theory.
If one knows an equation that permits deriving reliably, output values
from a list of inputs to the system under investigation,
then one has ``counterfactual definiteness'' (CFD) in the knowledge of that system~\cite{PEAR09}.

The word ``counterfactual'' is a misnomer~\cite{PEAR09} but is well established.
It is therefore helpful to have a clear-cut operational definition of what is meant with CFD.
In essence, CFD means that the output state of a system, represented by a vector of values $y$,
can be calculated using an explicit formula, e.g. $y = f(x)$
where $f(.)$ is a known vector-valued function of its argument $x$.
If $x$ denotes a vector of values, the elements of this vector must be
independent variables for the mathematical model to be CFD-compliant~\cite{HESS16b,RAED16c}.

In laboratory EPRB experiments, every trial takes place
under different conditions, different settings etc.  which may or may not affect the outcome of a single trial.
Therefore, data produced by laboratory EPRB experiments (or any other laboratory experiments)
can, as a matter of principle, not be cast in the form of data generated by a CFD-compliant model.
On the other hand, performing computer experiments in a CFD-compliant manner is not difficult nor is it much work
to change a CFD-compliant algorithm into one that does not meet all the requirements of CFD simulation.
In other words, computer experiments can be carried out in both CFD and non-CFD mode, providing
quantitative information about the differences of these two modes of modeling.

In the realm of finite sets of two-valued data,
a strict derivation of Bell-type inequalities~\cite{BELL93,BELL01}, such as the Bell-CHSH~\cite{CLAU69,BELL93}
and Eberhard's inequalities~\cite{EBER93} require
that these data are generated in a CFD-compliant manner~\cite{HESS16b,RAED16c}.
In other words, CFD is a prerequisite for deriving Bell-type inequalities.
Therefore, to test whether or not a simulation model produces data that do not satisfy such inequalities,
it is necessary to perform a CFD-compliant simulation.
Otherwise, there is no mathematical justification for the hypothesis
that these data should satisfy Bell-type inequalities in the first place.
Of course, we can always revert to the non-CFD algorithm and check if e.g. averages exhibit
the same features as the averages obtained from the CFD-compliant algorithm (see section~\ref{RES}).

In an earlier paper~\cite{RAED16c}, we have adopted this strategy to demonstrate that
in the case of EPRB experiments,
\begin{enumerate}
\item
CFD-compliant simulations can reproduce the averages and correlation of two particles in the singlet state,
\item
CFD does not distinguish classical from quantum physics because our computer models do not contain any quantum concepts,
yet yield results that lead to conclusions (e.g. entanglement) that are commonly regarded as signatures of quantum physics.
\end{enumerate}

In this paper, we adopt the same strategy.
We construct a CFD-compliant simulation model of the laboratory experiments~\cite{GIUS15,SHAL15}, meaning that
we simulate a perfected, idealized realization of these laboratory experiments.
Of course, this does not mean that we omit essential features of the laboratory experiments.
These features have to be included, otherwise the simulation model is not applicable to these laboratory experiments.
see section~\ref{MOD} for a general discussion of this aspect.

\section{Computational model of the laboratory experiments~\cite{GIUS15,SHAL15}}\label{COM}\label{LAB}

In this section, we introduce the essential elements of the simulation model of the laboratory experiments
reported in Ref.~\cite{GIUS15,SHAL15}. The details of the simulation algorithm are given in section~\ref{ALG}.
For concreteness, we adopt the terminology that is used in Ref.~\cite{GIUS15} when we connect
the elements of the simulation model to those of the laboratory experiments.

As shown in Fig.~\ref{fig0}, in a typical EPRB experiment there are three different components.
There is a source and there are two observation stations.
The algorithm that simulates the source is described in full detail in section~\ref{ALG}.
In this section, we focus on the observation stations which, because we are performing ``perfected'' experiments,
are assumed to be identical.

\begin{figure}[t]
\begin{center}
\includegraphics[width=0.3\textwidth]{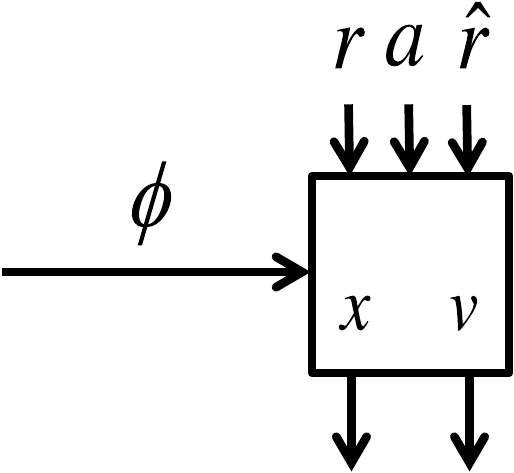}
\caption{%
Block diagram of an observation station. The specific form of the input-output relations
$x=x(a,\phi,r)$ and $v=v(a,\phi,\widehat r)$ is not important to follow
the discussion on the conceptual level and is only essential to perform
the simulations reported in this paper.
These relations are defined by Eqs.~(\ref{cfd0a}) and (\ref{cfd0b}) in section~\ref{ALG}.
}
\label{fig1}
\end{center}
\end{figure}

\subsection{Observation station}
In Fig.~\ref{fig1}, we show a graphical representation of the function of an observation station.
Input to an observation station is the setting $a$ (representing the orientation of the polarizer),
two numbers $0\le r,\widehat r<1$ taken from a list of uniform random numbers (see section~\ref{ALG} for further details)
and an angle $0\le \phi < 2\pi$ (representing the polarization of the photon).
Output of the observation station is a two-valued variable $x=\pm1$ and a detector-related variable $v_{\mathrm{min}}\le v\le v_{\mathrm{max}}$.

The correspondence between the data produced by the experimental realization of an observation station
and those generated by the computational model is as follows.
The variable $x$ encodes the detector outcomes (either $D_{+,i}$ or $D_{-,i}$ in station $i=1,2$) that fired.
In the laboratory experiments~\cite{GIUS15,SHAL15} there is only one detector in each station
but in the computer experiment we can easily simulate the complete EPRB experiment (see Fig.~\ref{fig0}), hence
we consider both the ``$+$'' and ``$-$'' events.
The variable $v$ represents the voltage signal produced by the electronics that amplifies the transition-edge detector current
(see the description in section IV of the supplementary material to Ref.~\cite{GIUS15}).

If necessary, we label different events by attaching
the subscript $i=1,2$ of the observation station and/or the subscript $k$ where $k=1,\ldots,N$
and $N$ denotes the total number of input events to a station.
In full detail, for the $k$th input at station $i$, the observation station $i$ generates
the output values $x_i=x_i(a_i,\phi_{i,k},r_{i,k})$ and $v_i=v_i(a_i,\phi_{i,k},\widehat r_{i,k})$
according to the rules which will be specified in full detail in section~\ref{RES}. 
Occasionally, we use the notation $x_i(a_i)=x_i(a_i,\phi_{i,k},r_{i,k})$
and $v_i(a_i)=v_i(a_i,\phi_{i,k},\widehat r_{i,k})$ to simplify the writing
while still emphasizing that the $x$'s and $v$'s only depend on variables
that are local to the respective station.

\subsection{Photon identification}\label{PHOTID}
In the following, we use the term {\sl detection event} whenever
the negative voltage signal produced by the electronics that amplifies the transition-edge detector current
is smaller (we are dealing with negative voltages) than the {\bf ``trigger threshold''}
(terminology from Ref.~\cite{GIUS15} (supplementary material)),
and speak of the observation of a {\sl photon} whenever the same negative voltage signal
is smaller than the {\bf ``photon identification threshold''} (about 4/3 times the ``trigger threshold'')
(terminology from Ref.~\cite{GIUS15} (supplementary material)).

From the description of the laboratory EPRB experiments under scrutiny, it follows immediately that
not every detection event is regarded as the observation of a photon~\cite{GIUS15,SHAL15}.
Indeed, after all the voltage traces of an experimental run have been recorded, a part of
the collected trace is analyzed by software, the photon identification thresholds are ``calibrated'' and
assuming that the relevant properties of the whole set of traces is stationary in real time,
the remaining set of traces is analyzed~\cite{GIUS15}.
In Ref.~\cite{GIUS15} there is no specification of the cost function that is being minimized
by the calibration procedure whereas Ref.~\cite{SHAL15}(supplementary material) explicitly states that
{\it ``Because the experiment was calibrated to maximize violation of the CH inequality...''}.
This seems to suggest that the software is designed to adjust the photon identification thresholds such that
the desired result, namely a violation of a Bell-type inequality, is obtained.

In our simulation approach, we may assume that all units are identical.
Therefore, unlike in Ref.~\cite{GIUS15}, one and the same value of photon identification threshold, denoted by ${\cal V}$,
can be used to identify photons.
The effect of the photon identification threshold is captured by the function
\begin{eqnarray}
w(a_i)&=&w_i(a_i,\phi_{i,k},\widehat r_{i,k})=\Theta({\cal V}-v_i)\quad,\quad i=1,2
,
\label{cfd2x}
\end{eqnarray}
where $\Theta(x)$ is equal to one if $x>0$ and is zero otherwise.
Recall that, as in Ref.~\cite{GIUS15}, ${\cal V}$ is negative.
In the simulation, we do not ``calibrate'' ${\cal V}$ but simply generate the data and analyze the results
as a function ${\cal V}$.

The correspondence with the data collected in the laboratory experiment is as follows:
a detection event is represented by $x_i(a_i)=+1$ and $w_i(a_i)=0$ and
the observation of a photon in station $i=1,2$
is represented by $x_i(a_i)=+1$ (because there is only one, not two, transition-edge detectors at each station)
and $w_i(a_i)=+1$ (implemented in software), both exactly as in the simulation model.
Recall, and this is new and important, that also in the simulations the photon identification is performed locally,
i.e. without communication between the observation stations.

\section{Equivalence of local time-window and time-coincidence processing}\label{EQU}

In this section we show that in spite of the conceptually very different setup,
from an operational point of view,
employing local photon identification thresholds is equivalent to local time-window selection and also to time-coincidence counting
that is used in most EPRB experiments with photons~\cite{KOCH67,CLAU78,ASPE82b,WEIH98}.

As explained above, in the laboratory experiments a detection event is classified as being
a photon if the (negative) voltage signal, denoted by $v$, produced by the electronics that amplifies the transition-edge detector current
(see the description in section IV of the supplementary material to Ref.~\cite{GIUS15})
is smaller than the photon identification threshold ${\cal V}$.
This rejection criterion is implemented through Eq.~(\ref{cfd2x}) from which it follows directly
that the criterion to observe a photon in these laboratory experiments is $v\le {\cal V}$.
Recall that we adopted the convention of the laboratory experiments~\cite{GIUS15} in which ${\cal V}$ takes negative values.

In practice, we have $v_\mathrm{min}\le v_i \le v_\mathrm{max}$ and $v_\mathrm{min}\le {\cal V} \le v_\mathrm{max}$
with finite $v_\mathrm{min}$ and $v_\mathrm{max}$,
hence the condition for counting a detection event as photon may be written as
\begin{eqnarray}
0\le \frac{v_i-v_\mathrm{min}}{v_\mathrm{max}-v_\mathrm{min}}\le
\frac{{\cal V}-v_\mathrm{min}}{v_\mathrm{max}-v_\mathrm{min}}\quad,\quad i=1,2
.
\label{ear0}
\end{eqnarray}
Defining a dimensionless ``time'' $t_i\equiv (v_i-v_\mathrm{min})/(v_\mathrm{max}-v_\mathrm{min})$
and a dimensionless ``time window'' $W=({\cal V}-v_\mathrm{min})/(v_\mathrm{max}-v_\mathrm{min})$,
Eq.~(\ref{ear0}) reads
\begin{eqnarray}
0\le t_i \le W\quad,\quad i=1,2
,
\label{ear1}
\end{eqnarray}
which expresses the condition to observe a photon at station $i=1,2$
in terms of {\bf locally} defined time slots of size $W$.
From Eq.~(\ref{ear1}) we have $-t_2 \le t_1-t_2 \le W - t_2$ and using $-W \le -t_2$  we find
\begin{eqnarray}
|t_1-t_2| \le W
,
\label{ear2}
\end{eqnarray}
which is exactly the same criterion as the one used
in most EPRB experiments with photons~\cite{KOCH67,CLAU78,ASPE82b,WEIH98}
and in computer simulation models thereof~\cite{PASC86,RAED06c,RAED07a,RAED07b,RAED07c,ZHAO08,RAED12,RAED13a,RAED16c}.

In summary: although physically very different, local voltage thresholds,
local time windows or time-coincidence counting are mathematically equivalent and
all serve the same purpose, namely to give an operational meaning
to the statement ``a single photon (pair)'' has been identified.

\section{Loopholes in experimental tests of Bell's theorem}\label{MOD}

A useful physical theory of an experiment needs to encompass all relevant parameters that affect the experimental outcomes
and, of course, the most important elements of physical reality, namely the data itself.
Specifically, a physical theory that describes pair-correlations of space-like separated systems,
must account for and include all procedures that determine the detection of the particles
and the knowledge which pair of particles and data belongs together.
Therefore, any model which purports to describe the laboratory experiments~\cite{GIUS15,SHAL15} that we consider
in this paper must necessarily account for the photon identification threshold mechanism
that is instrumental in the data-processing step of these experiments, see section~\ref{PHOTID}.
Likewise, the earlier generation of EPRB experiments that employ time coincidence
to identify pairs~\cite{CLAU74,ASPE82b,WEIH98} can only be faithfully be described by models
that incorporate the time-coincidence window selection process that is an essential component of
this class of experiments~\cite{PASC86,RAED06c,RAED07a,RAED07b,RAED07c,ZHAO08,RAED12,RAED13a,RAED16c}.

Drawing a conclusion about a world view from models (such as those of Bell and his followers, see the Appendix)
that do not properly account for the photon identification threshold mechanism which,
in the laboratory experiments~\cite{GIUS15,SHAL15}, is essential for identifying the photons,
requires a drastic departure from rational reasoning.
If we allow for such a departure, we might equally well wonder what it means for our wold view
when we construct and analyze a model of an airplane that excludes the engines
and then observe that a real airplane can take off by itself.
Any reasonable person would rightfully question our ability to represent the airplane (or laboratory experiments) by such a model
and regard the idea that we may have to change our world because of the contradictions to such a model as unfounded.
In other words, the only logically correct conclusion that one can draw from the failure of Bell-type models
to describe the qualitative features of the experimental data is that these models are too simple,
which in this case is obvious as they miss at least one important ingredient: the photon identification mechanism.

The photon-identification loophole that we introduce in this paper accounts for
\begin{enumerate}[(i)]
\item
the fact that laboratory experiments~\cite{GIUS15,SHAL15} employ a threshold to decide whether or not a detection
event is considered to be a photon,
\item
the assumption that voltage signals produced by the detection equipment may depend on the analyzer setting (see Eq.~(\ref{cfd2x})).
Regarding this latter assumption, it is of interest to recall that since the early days of the Bell-test experiments,
it is well-known that application of Bell-type models requires at least one extra assumption.
We reproduce here the relevant passage from Ref.~\cite{CLAU74} (p.1890):
{\it ``The approach used by CHSH is to introduce an auxiliary assumption, that if a particle passes through a spin
analyser, its probability of detection is independent of the analyser's orientation.
Unfortunately, this assumption is not contained in the hypotheses of locality, realism or determinism.''}
\item
the requirement that a relevant model of an experiment needs to encompass all elements that affect the experimental outcomes.
\end{enumerate}

There is a large body of theoretical work that considers all kinds of loopholes in experiments that must be closed before a definite
conclusion about the consequences of Bell inequality tests for certain wold views can be drawn.
A detailed, comprehensive discussion of a large collection of loopholes is given in Ref.~\onlinecite{LARS14}.
Also in this respect, the digital computer -- laboratory experiment metaphor
offers unique possibilities because we can open and close loopholes at will.
As this and our earlier paper~\cite{RAED16c} demonstrate, computer simulation models of EPRB experiments
can easily be engineered to be free of e.g. detection, coincidence, and memory loopholes~\cite{LARS14}
and, in addition, include features such as CFD compliance that close the contextuality loophole~\cite{NIEU11}.

Wrapping up: in this paper we construct a minimal model of the perfected version of the laboratory experiment~\cite{GIUS15}.
With the exception of the photon-identification loophole, this minimal model is free of the known loopholes
and reproduces the quantum results of the EPRB thought experiment, from which violations follow automatically.
This approach offers the unique possibility to confront all kinds of reasonings and assumptions, such as the (ii) above,
with actual facts.

\section{CFD compliance}\label{CFD}

In section~\ref{COM}, the operation of the simulation model of an observation station
has been defined such that for every input event $(a,\phi,r,\widehat r)$, we know the values
of all outputs variables $x=x(a,\phi,r)$ and $v=v(a,\phi,\widehat r)$.
Therefore, the input-output relation of this unit,
represented by the diagram of Fig.~\ref{fig1}, satisfies the requirement of a CFD-compliant model.

\begin{figure}[t]
\begin{center}
\includegraphics[width=0.7\textwidth]{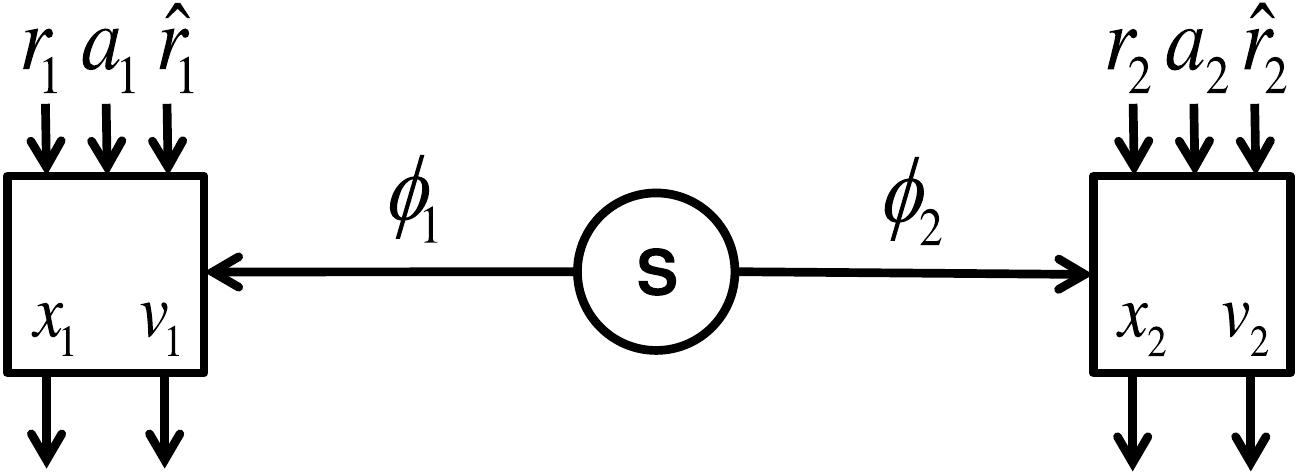}
\caption{Schematic layout of the computational equivalent of the laboratory EPRB experiments reported
in Refs.~\cite{GIUS15,SHAL15}.
The input-output relation for $i=1,2$ is given by
$x_i=x_i(a_i,\phi_i,r_i)$ and $v_i=v_i(a_i,\phi_i,\widehat r_i)$ defined by Eqs.~(\ref{cfd0a}) and (\ref{cfd0b}).
Alternatively, the input-output relation may be written as
$(x_1,v_1,x_2,v_2)=F(a_1,a_2,\phi_1,\phi_2,r_1,r_2,\widehat r_1,\widehat r_2)$
showing that {\bf for fixed $(a_1,a_2)$} the simulation model satisfies the definition of a CFD theory~\cite{HESS16b}.
}
\label{fig2}
\end{center}
\end{figure}

The computational equivalent of the EPRB experiments~\cite{GIUS15,SHAL15} is shown in Fig.~\ref{fig2}.
Each time the source $\mathbf{S}$ is activated, it sends one entity carrying the data $\phi_1$
to station 1 and another entity carrying the data $\phi_2$ to station 2.
The procedure for generating the $\phi$'s, $r$'s and $\widehat r$'s is specified in section~\ref{RES}.

\begin{figure}[t]
\begin{center}
\includegraphics[width=0.9\textwidth]{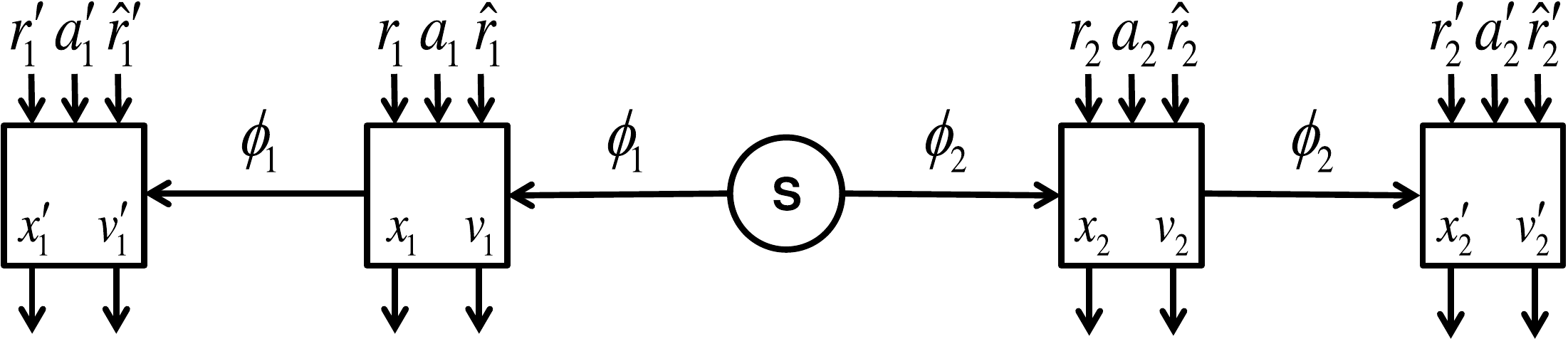}
\caption{Computational model for the EPRB experiment satisfying the criterion of a CFD theory.
}
\label{fig3}
\end{center}
\end{figure}

Upon arrival of the entities, observation stations $i=1,2$ execute their internal algorithm
(completely specified by Eqs.~(\ref{cfd0a}) and (\ref{cfd0b}))
and produces output in the form of the pair $(x_i,v_i)$.
The scheme represented by Fig.~\ref{fig2} computes the vector-valued function
\begin{eqnarray}
\left(\begin{array}{r}
        x_1 \\
        v_1\\
        x_2\\
        v_2
\end{array}\right)
&=&
\left(\begin{array}{c}
x_1=x_1(a_1,\phi_1,r_1)\\
v_1=v_1(a_1,\phi_1,\widehat r_1)\\
x_2=x_2(a_2,\phi_2,r_2)\\
v_2=v_2(a_2,\phi_2,\widehat r_2)\\
\end{array}\right)
=\mathbf{F}(a_1,a_2,\phi_1,\phi_2,r_1,r_2,\widehat r_1,\widehat r_2)
,
\label{cfd1}
\end{eqnarray}
which clearly defines a CFD-compliant model.
Nevertheless, with this CFD-compliant model we {\bf cannot} construct the quadruple $(x_1,x_2,x_1^\prime,x_2^\prime)$ in a CFD-compliant manner.
Indeed, by construction, there is no guarantee that the
$(\phi_{i,k},r_{i,k})$'s that determine say the $x_1$'s for the pair of settings $(a_1,a_2)$
will be the same as the $(\phi_{i,k},r_{i,k})$'s that determine
that values of the $x_1^\prime$'s for the pair of settings $(a_1^\prime,a_2)$.
Of course, with a simulation on a digital computer being an ideal, fully controllable experiment,
we could enforce CFD-compliance by re-using the same $(\phi_{i,k},r_{i,k},\widehat r_{i,k})$'s
for every pair of settings. This would make the simulation CFD-compliant.
However, in this paper we do not so  but instead generate new values of the $(\phi_{i,k},r_{i,k},\widehat r_{i,k})$'s
for every new instance of input.

The layout of a CFD-compliant computer model of the EPRB experiment is depicted in Fig.~\ref{fig3}.
It uses the same units as the model shown in Fig.~\ref{fig2}, the only
difference being that the input $\phi_i$ is now fed into
an observation station with setting $a_i$ and into another one with setting $a_i^\prime$,
something which, for obvious reasons, is impossible to realize in laboratory experiments with photons.
As each of the four units operates according to the rules given by Eq.~(\ref{cfd0a}) and (\ref{cfd0b}), we have
$(x_1,x_1^\prime,x_2,x_2^\prime)=\mathbf{X}(a_1,a_1^\prime,a_2,a_2^\prime,\phi_1,\phi_2,r_1,r_1^\prime,r_2,r_2^\prime)$
and
$(v_1,v_1^{\prime},v_2,v_2^{\prime})=\mathbf{T}(a_1,a_1^\prime,a_2,a_2^\prime,\phi_1,\phi_2,\widehat r_1,\widehat r_1^\prime,\widehat r_2,\widehat r_2^\prime)$.
As the arguments of the functions $\mathbf{X}$ and $\mathbf{T}$ are independent and may take any
value out of their respective domain, the whole system represented by Fig.~\ref{fig3}
satisfies, by construction, the criterion of a CFD theory.

Note that the actual EPRB experiments produce only pairs of data. The three pairs of data considered by Bell involve, therefore,
six local measurements and the four pairs of CHSH involve eight local measurements.
Our CFD compliant model considers only quadruple (= four local) measurements
to simulate the actual eight possible measurement outcomes of a CHSH type experiment.

\section{Bell-type inequalities}\label{BEL}

It is evident from the formulation of his model that Bell and all his followers, including Wigner,
do not deal with the issue of identifying particles and take for granted that the measured pairs correspond
to the correlated particles.
The common prejudice that additional variables cannot possibly defeat Bell-type inequalities is based on the assumption that all
sent out correlated pairs, or a representative sample of them, are measured.
This reasoning does not account for the photon identification or
pair-modeling loophole: the necessary particle or pair identification may necessarily select in a way that is not
representative for all possible measurements of all possible pairs emanating from the source.
In this section, we adopt Bell's viewpoint by ignoring the $v$-variables and
demonstrate that CFD-compliance and the existence of Bell-type inequalities
are mathematically equivalent.

Figure.~\ref{fig3} shows the CFD compliant arrangement of the computer experiment. The two stations
on the left of the source {\bf S} receive the same data $\phi_1$ from the source.
The settings $a_1$ and $a_{1}^\prime$ are fixed for the duration of the $N$ repetitions of the experiment.
The same holds for the two stations on the right of the source, with subscript 1 replaced by 2.
Clearly, the algorithm represented by Fig.~\ref{fig3} generates quadruples of output data
$(x_1^{\phantom{\prime}},x_1^\prime,x_2^{\phantom{\prime}},x_2^\prime)$ in a CFD-compliant manner.

For any such quadruple $(x_1^{\phantom{\prime}},x_1^\prime,x_2^{\phantom{\prime}},x_2^\prime)$ in which the $x$'s only take
values +1 and -1, it is straightforward to verify that the following equalities hold:
\begin{eqnarray}
b_1&=& x_1 x_1^\prime + x_1 x_2 + x_1^\prime x_2
=\left\{\begin{array}{r} -1\\ +3 \end{array}\right.
\label{cfd1a1}
\\
b_2&=& x_1 x_1^\prime + x_1 x_2^\prime + x_1^\prime x_2^\prime
=\left\{\begin{array}{r} -1\\ +3 \end{array}\right.
\label{cfd1a2}
\\
b_3&=& x_1 x_2 + x_1 x_2^\prime + x_2 x_2^\prime
=\left\{\begin{array}{r} -1\\ +3 \end{array}\right.
\label{cfd1a3}
\\
b_4&=& x_1^\prime x_2 + x_1^\prime x_2^\prime + x_2 x_2^\prime
=\left\{\begin{array}{r} -1\\ +3 \end{array}\right.
\label{cfd1a4}
\\
s&=& x_1 x_2 - x_1 x_2^\prime + x_1^\prime x_2 + x_1^\prime x_2^\prime
=\left\{\begin{array}{r} -2\\ +2 \end{array}\right.
.
\label{cfd1a}
\end{eqnarray}
Other equivalent sets of equalities can be obtained by replacing e.g. $x_1$ by $-x_1$ etc.
Note that e.g. Eq.~(\ref{cfd1a1}) follows from Eq.~(\ref{cfd1a}) if we set $x_2^\prime=x_1$.

In a non-CFD setting, the data is collected as four pairs which we may denote as
$(x_1^{\phantom{\prime}},x_2^{\phantom{\prime}})$,
$(\widetilde x_1^{\phantom{\prime}},x_2^{\phantom{\prime}})$,
$(x_1^\prime,\widetilde x_2^{\phantom{\prime}})$, and
$(\widetilde x_1^\prime,\widetilde x_2^\prime)$
where the tilde is used to indicate that the value of e.g. $x_1^{\phantom{\prime}}$
obtained with setting $(a_1,a_2^\prime)$ may be different from the one
obtained with setting $(a_1,a_2)$.
Instead of Eq.~(\ref{cfd1a}), we now consider the expression
$\widetilde s= x_1 x_2 - \widetilde x_1 x_2^\prime + x_1^\prime \widetilde x_2 +
\widetilde x_1^\prime \widetilde x_2^\prime=-4,-2,0,+2,+4$
and similar ones for $\widetilde b_1,\ldots,\widetilde b_4$, each of them taking values $-3,-1,+1,+3$.
If we now impose that $\widetilde s=-2,+2$ and $\widetilde b_1\ldots,\widetilde b_4=-1,3$,
simple enumeration of all possible values of the $x$'s and the $\widetilde x$'s shows that in order
for all equalities to be satisfied simultaneously we must have
$\widetilde x_1^{\phantom{\prime}}=x_1^{\phantom{\prime}}$,
$\widetilde x_1^{\prime}=x_1^{\prime}$,
$\widetilde x_2^{\phantom{\prime}}=x_2^{\phantom{\prime}}$,
$\widetilde x_2^{\prime}=x_2^{\prime}$.
In other words, imposing the constraints
$\widetilde s=-2,+2$ and $\widetilde b_1=-1,3,\ldots,\widetilde b_4=-1,3$
on data obtained in a non-CFD setting
forces this data to form quadruples, i.e. to be CFD compliant.
It then follows immediately that CFD is {\bf necessary and sufficient} for
the equalities Eqs.~(\ref{cfd1a1}) -- (\ref{cfd1a}) to hold.

Attaching the subscript $k$ ($k=1,\ldots,N$) to label the events,
the algorithm generates the set of quadruples
$\{(x_{1,k}^{\phantom{\prime}},x_{1,k}^\prime,x_{2,k}^{\phantom{\prime}},x_{2,k}^\prime)\;|\;k=1,\ldots,N\}$.
Introducing the Bell-CHSH function
\begin{eqnarray}
\widehat S&=&\frac{1}{N}\sum_{k=1}^N s_k
,
\label{cfd1b}
\end{eqnarray}
it follows immediately from $|s_k|=2$ (see Eq.~(\ref{cfd1a})) that $|\widehat S|\le2$ for all $N\ge1$, that is we obtain the
Bell-CHSH inequality constraining four correlations of pairs of actual data.
Put differently, if the output consists of quadruples of two-valued data generated by the setup shown in Fig.~\ref{fig3}
and we ignore the $v$-variables then the Bell-CHSH inequality $|\widehat S|\le2$ is always satisfied,
independent of the number of events $N\ge1$ considered.

Similarly, from the fact that for example $b_{1,k}=-1,3$ we obtain the Leggett-Garg inequality~\cite{LEGG85,HESS16a} for
three correlations of pairs of actual data
and by combining $b_{1,k}=-1,3$ with the equalities obtained by substituting $x_1\rightarrow -x_1$
we obtain the Bell inequality involving three correlations of pairs of actual data~\cite{BELL93}.
In other words, Bell-type inequalities follow directly from the fact that quadruples of data satisfy
rather trivial arithmetic identities such as Eq.~(\ref{cfd1a}).

It then also follows immediately that CFD is a {\bf necessary and sufficient} condition for
the data
$(x_{1,k}^{\phantom{\prime}},x_{2,k}^{\phantom{\prime}})$,
$(\widetilde x_{1,k}^{\phantom{\prime}},x_{2,k}^{\phantom{\prime}})$,
$(x_{1,k}^\prime,\widetilde x_{2,k}^{\phantom{\prime}})$, and
$(\widetilde x_{1,k}^\prime,\widetilde x_{2,k}^\prime)$ with $k=1,\ldots,N$
to satisfy simultaneously for all $N\ge1$, all Bell-type inequalities involving three and four different correlations of pairs.
We emphasize that this conclusion follows from elementary arithmetic only.
Concepts such as ``locality'' or any other physical argument are irrelevant for establishing this result.

Similar reasoning yields Eberhard's inequality
which differs from the Bell-CHSH  inequality in the sense that it can account for reduced detector efficiencies~\cite{EBER93}.
For convenience of comparison with the original work, we temporarily adopt Eberhard's parlance and notation.
Central to Eberhard's derivation is the so-called {\sl fate} of a photon.
This fate can be either detected in the ordinary beam (labeled $o$), or
detected in the extraordinary beam (labeled $e$), or undetected (labeled $u$).
For counting purposes, we represent the fate of a photon by the symbol $f$, taking the
values $+1$, $0$, and $-1$ corresponding to $o$, $u$ and $e$, respectively.
Introducing the variables $n_o=f(f+1)/2$, $n_e=f(f-1)/2$, and $n_u=(1-f^2)$,
it is clear that one of them takes the value $1$ with the other two taking the value $0$.
For a given pair of settings, say $(\alpha_1,\beta_2)$, the number of pairs with both photons suffering fate $(o)$
is then given by $n_{oo}(\alpha_1,\beta_1)=n_{o}(\alpha_1)n_{o}(\beta_1)=f_{1,1}(f_{1,1}+1)f_{2,1}(f_{2,1}+1)/4$
where $f_{1,1}=f_{1,i}(\alpha_i)$ and $f_{2,i}=f_{2,i}(\beta_i)$ for $i=1,2$.
There are similar expressions for $n_{eo}(\alpha_1,\beta_2)$, $n_{uo}(\alpha_1,\beta_2)$, etc.
Following Eberhard, we consider the expression~\cite{EBER93}
\begin{eqnarray}
j&=& n_{oe}(\alpha_1,\beta_2)+n_{ou}(\alpha_1,\beta_2)+n_{eo}(\alpha_2,\beta_1)
\nonumber \\
&&+n_{uo}(\alpha_2,\beta_1) +n_{oo}(\alpha_2,\beta_2)-n_{oo}(\alpha_1,\beta_1)
.
\label{cfd2a}
\end{eqnarray}
It is straightforward to enumerate all possible $81$ values of the 4 different $f$-variables that appear in Eq.~(\ref{cfd2a}).
This enumeration proves that $j\ge0$, independent of the values of the settings.
Attaching the subscript $k$ ($k=1,\ldots,N$) to label the events as we did
to derive the Bell-CHSH inequality and introducing the Eberhard function
\begin{eqnarray}
J_\mathrm{Eberhard}&=&\sum_{k=1}^N j_k
,
\label{cfd2b}
\end{eqnarray}
it follows immediately that $J_\mathrm{Eberhard}\ge0$ for all $N\ge1$.

In the laboratory experiments~\cite{GIUS15,SHAL15} there is only one detector per observation station.
Hence it makes sense to regard also say, the $e$ photons, as undetected.
In terms of the ``fate'' variables $f$ introduced above
this amounts to letting $f$ taking the values $+1$ and $0$ corresponding to $o$ and $u$, respectively.
Instead of Eq.~(\ref{cfd2a}), we now consider the expression
\begin{eqnarray}
j_\mathrm{CH}&=&n_{ou}(\alpha_1,\beta_2)+n_{uo}(\alpha_2,\beta_1) +n_{oo}(\alpha_2,\beta_2)-n_{oo}(\alpha_1,\beta_1)
.
\label{cfd2c}
\end{eqnarray}
Enumerating all possible $16$ values of the 4 different $f$-variables that appear in Eq.~(\ref{cfd2b})
proves that $j_\mathrm{CH}\ge0$, independent of the values of the settings.
Attaching the subscript $k$ ($k=1,\ldots,N$) to label the events as before
and introducing the CH function
\begin{eqnarray}
J_\mathrm{CH}&=&\sum_{k=1}^N j_{\mathrm{CH},k}
,
\label{cfd2d}
\end{eqnarray}
it follows immediately that the CH inequality~\cite{CLAU74} $J_\mathrm{CH}\ge0$ holds for all $N\ge1$.

In short: if for all $N$, the $x$'s ($f$'s) are generated according to a CFD-compliant procedure
the Bell-CHSH (the Eberhard and CH) inequality is (are) satisfied.
In essence, this result is embodied in the work of George Boole~\cite{BO1862}, see also Ref.~\cite{PITO94,RAED11a}.
Moreover, as CFD implies that all Bell-type inequalities hold for all $N\ge1$,
there is no room for speculating without violating at least one of the rules of Aristotelian logic that
something ``spooky'' is going on if we encounter data that violate a Bell-type inequality.
The logically correct conclusion that one can draw from such a violation
is that these data have not been generated in a CFD-compliant manner.

\section{An inequality accounting for photon identification}\label{INE}

In this section, we address the modifications to the inequality $|\widehat S|\le2$ that ensue when
we take into account the fact that laboratory experiments
employ the photon identification threshold to decide whether or not a detection event corresponds
to the observation of a photon.

The average {\bf detection event} counts and {\bf detection event} pair correlation are given by
\begin{eqnarray}
\widehat E_i(a_i)&=&\frac{1}{N}\sum x_i(a_i)\quad,\quad i=1,2
\nonumber \\
\widehat E(a_1,a_2)&=&\frac{1}{N}\sum x_1(a_1)x_2(a_2)
,
\label{cfd3a}
\end{eqnarray}
respectively, and we have similar expressions for the other choices of settings.
In Eq.~(\ref{cfd3a}) and in the equations that follow, it is understood that
$\sum$ means $\sum_{k=1}^N$, i.e. the sum over all input events,
represented by values of the $\phi$'s.
As shown in section~\ref{BEL}, if the $x$'s that enter Eq.~(\ref{cfd3a}) have been obtained by a CFD-compliant procedure,
the correlations $\widehat E(a_1,a_2),\ldots$ satisfy Bell-type inequalities.

In contrast to Eq.~(\ref{cfd3a}), the average {\bf photon} counts and {\bf photon} pair correlation
for the settings $(a_1,a_2)$ are given by
\begin{eqnarray}
E_i(a_i)&=&\frac{\sum w(a_i)x_i(a_i)}{\sum w(a_i)}\quad,\quad i=1,2
\nonumber \\
E(a_1,a_2)&=&\frac{\sum w(a_1)w(a_2)x_1(a_1)x_2(a_2)}{\sum w(a_1)w(a_2)}
,
\label{cfd3}
\end{eqnarray}
where, as explained in section~\ref{PHOTID}, the $w$'s in Eq.~(\ref{cfd3})
account for the effect of the photon identification thresholds and take values 0 or 1.
Clearly, Eq.~(\ref{cfd3}) is very different from Eq.~(\ref{cfd3a})
unless all the $w$'s that appear in Eq.~(\ref{cfd3}) are equal to 1,
in which case the photon identification threshold mechanism is superfluous and
unlike as in the laboratory experiment~\cite{GIUS15},
the number of photon and detection events is the same.

In the analysis of the experimental data, the photon identification threshold is
chosen such that many of the $w$'s are zero~\cite{GIUS15}.
Hence from the discussion in section~\ref{BEL}, it follows immediately that with some $w$'s zero,
it is impossible to prove that the Bell-CHSH function
$ S= S(a_1,a_2,a_1^\prime,a_2^\prime)
\equiv  E(a_1,a_2)- E(a_1,a_2^\prime)+  E(a_1^\prime,a_2)+ E(a_1^\prime,a_2^\prime)$
satisfies the inequality $|S|\le2$.

However, it directly follows from the proof
given in our earlier paper~\cite{RAED16c} that if the $x$'s and $w$'s have been generated
by a CFD-compliant procedure, the Bell-CHSH function $S$ can never violate the inequality
\begin{equation}
|S|=\left|
E(a_1,a_2)-E(a_1,a_2^\prime)+E(a_1^\prime,a_2)+E(a_1^\prime,a_2^\prime)
\right|\le 4-2\delta
\
.
\label{cfd11}
\end{equation}
The term $2\delta$ in Eq.~(\ref{cfd11}) is a measure for the number of paired events that have been
rejected relative to the number of emitted pairs.
In detail, $0\le \delta\equiv N'/N_{\mathrm{max}}\le 1$ where $N^\prime$
denotes the number of input events for which the negative voltage signal of all the photons
is smaller than the photon identification threshold ${\cal V}$ and
$N_{\mathrm{max}}$ is the maximum number of contributing pairs per setting.
If all paired events would be regarded as photon pairs then $\delta=1$
and then, and only then we recover the Bell-CHSH inequality $|S|\le2$.
If the $x$'s and $w$'s have not been generated by a CFD-compliant procedure, there is only the trivial bound $|S|\le4$.

The inequality Eq.~(\ref{cfd11}) is a rigorous mathematical fact that holds if,
for all $N\ge1$, the $x$'s and $v$'s are generated in a CFD-compliant manner
and none of the denominators in Eq.~(\ref{cfd3}) is identically zero (in which case no photon pairs have been detected).
Conversely, if we find a set of $x$'s and $v$'s that yields a value of $|S|$ that exceeds
$4-2\delta$, we can only conclude that these data have not been obtained from a CFD-compliant procedure.
Any other conclusion would not be logically justified.

In analogy with the derivation of Eq.~(\ref{cfd11}),
one may derive an Eberhard-type or CH-type inequality that accounts for the $w$'s
but as such inequalities do not add anything to the discussion that follows, we do not discuss them any further.

\section{Discrete-event simulation algorithm}\label{ALG}

In this section, we specify the algorithm and the simulation procedure in full detail.
The algorithm that mimics the operation of the particle source is very simple.
For each event $k=1,\ldots,N$, a uniform random generator is used
to generate a floating-point number $0\le\phi_{1,k}\le 2\pi$.
This number is input to the stations with setting $(a_1,a_1^\prime)$
and another number $\phi_{2,k}=\phi_{1,k}+\pi/2$ is input to the stations with setting $(a_2,a_2^\prime)$.
Because of $\phi_{2,k}=\phi_{1,k}+\pi/2$, the $k$th event simulates the emission of a photon pair
with maximally correlated, orthogonal polarizations.
In this respect, we deviate from what is done in the laboratory experiments~\cite{GIUS15,SHAL15} in the following sense.
Unlike in the computer simulation, the detectors used in these laboratory experiments are not perfect.
As already mentioned, Eberhard's inequality can account for reduced detector efficiencies and this
feature can be put to good use through minimizing the value of $J_\mathrm{Eberhard}$ with respect to the correlation~\cite{EBER93}.
This is what is done in the laboratory experiments~\cite{GIUS15,SHAL15}.
However, in our simulation model, the detectors are perfect.
Hence the minimum value of $J_\mathrm{Eberhard}$
will be obtained by choosing maximally correlated, orthogonal polarizations~\cite{EBER93}.
Recall that our aim here is to simulate the most ideal, perfect experiment that accounts for all
the essential features of the laboratory experiments,
not to simulate a real laboratory experiment including trams passing by~\cite{GIUS15}  etc.

Upon receiving the input $\phi$ an observation station (see Fig.~\ref{fig1})
executes the following two steps.
First it retrieves two uniform random numbers $r$ and $\widehat r$ from a list of such numbers
(or, more conveniently, generates these numbers on the fly) and then
\begin{eqnarray}
&1.\;&\mathrm{computes\ \ \ }
c=\cos[2(a-\phi)]\;,\; s=\sin[2(a-\phi)]
\label{cfd0a}
\\
&2.\;&\mathrm{sets\ \ \ }
x=\sign(1+c-2*r) \;,\; v=\widehat r |s|^d (V_{\mathrm{max}}-V_{\mathrm{min}})-V_{\mathrm{max}}
,
\label{cfd0b}
\end{eqnarray}
where $d$, is an adjustable parameter to be discussed later
and $0\le V_{\mathrm{max}}$, and $0\le V_{\mathrm{min}}\le V_{\mathrm{max}}$ set the range of the voltage signal.
From Eq.~(\ref{cfd0b}) it follows that  $-V_{\mathrm{max}}\le v,{\cal V} \le -V_{\mathrm{min}}$,
as in the laboratory experiment~\cite{GIUS15}.

Our choice for the specific functional forms of $x=x(a,\phi,r)$ and $v=v(a,\phi,\widehat r)$ is inspired
by previous work in which it was shown that a similar model, which employs time-coincidence to identify pairs,
exactly reproduces the single particle averages and two-particle correlations
of the singlet state if the number of events becomes very large~\cite{RAED06c,MICH14a}.

Equations~(\ref{cfd0a}) and (\ref{cfd0b}) form the core of the simulation algorithm which
has the following key features:
\begin{itemize}
\item
For any fixed value of $\phi$ and uniformly distributed random numbers $r$,
the unit generates a sequence of randomly distributed $x$'s such
that the average of the $x$'s agrees (within statistical fluctuations) with Malus' law, i.e. the normalized frequencies
to observe $x=+1$ and $x=-1$ are given by $\cos^2(a-\phi)$ and $\sin^2(a-\phi)$, respectively.
\item
The presence of an output variable $v$ which serves to mimic the detector traces recorded in the laboratory experiments.
Note that the explicit expression of $v=v(a,\phi,\widehat r)$ shows a dependence on the local setting of the station.
Such a dependence cannot be ruled out a posteriori 
but finds a post-factum justification in the fact that the simulations reproduce the results for two particles in a singlet state,
see section~\ref{POS}.
\item
The use of the random numbers $0 \leq r,\widehat r \le 1$ mimics the uncertainties
about the outcomes, as observed in experiments.
Thereby, it is implicitly understood that for every instance
of new input, new values of the uniform random numbers $r$ and $\widehat r$ have been generated.
\item
By construction, the algorithm is a metaphor for Einstein-local experiments:
changing $a_1$ ($a_2$) does not affect the present, past or future values of $x_2$ ($x_1$) or $v_2$ ($v_1$).
In plain words, the output of one particular unit depends on the input to that particular unit only.
\end{itemize}

For the settings of the observation stations, we take
$a_1=\theta+\pi/8$, $a_1^\prime=a_1+\pi/4$, $a_2=\pi/8$, $a_2^\prime=3\pi/8$ and
let $\theta$ vary from 0 to $\pi$.
For this choice of settings, quantum theory for a system in the singlet state predicts
$E(a_1,a_2)=E(a_1^\prime,a_2^\prime)=\cos 2\theta$,
$E(a_1,a_2^\prime)=E(a_1^\prime,a_2)=\sin 2\theta$
and $S(a_1,a_2,a_1^\prime,a_2^\prime)=-2\sqrt{2}\cos(2\theta+\pi/4)$,
the latter reaching its maximum $2\sqrt{2}$ at $\theta=3\pi/8$.
When we operate the computer model in non-CFD mode, random numbers are
used to make a choice between the settings $a_i$ and $a_i^\prime$, for $i=1,2$, exactly
as in the experiments~\cite{GIUS15,SHAL15}.

The simulation procedure is quite simple. We choose a fixed photon identification threshold ${\cal V}$,
generate input pairs $k=1,\ldots,N$, collect corresponding outputs in terms of $x$'s and $w$'s,
and compute the single- and two-particle averages according to Eq.~(\ref{cfd11}), the Bell-CHSH function
$S(a_1,a_2,a_1^\prime,a_2^\prime)=E(a_1,a_2)-E(a_1,a_2^\prime)+E(a_1^\prime,a_2)+E(a_1^\prime,a_2^\prime)$,
and the Eberhard function $J$ given by Eq.~(\ref{cfd2b}) .

Because the computer experiment is ``perfect'', it differs from
the laboratory experiment in the sense that all pairs are created ``on demand''
and all emitted pairs create one detection event in each station (there are no ``false'' detection events)
but exactly as in the laboratory experiment, the local photon identification threshold
at each observation station serves to decide whether a photon has arrived or not.

\section{Computer simulation results}\label{RES}

\begin{figure}[t]
\begin{center}
\includegraphics[width=0.45\textwidth]{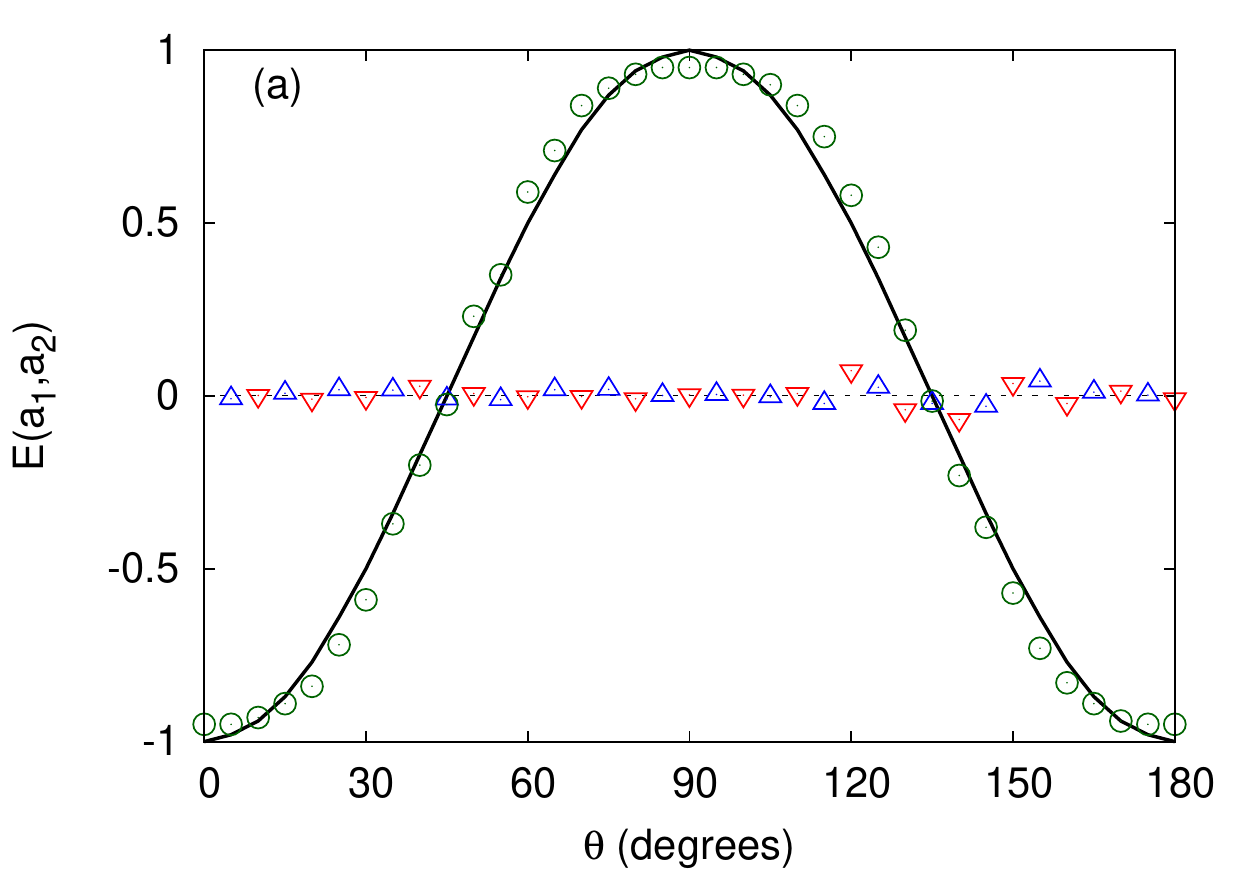}
\includegraphics[width=0.45\textwidth]{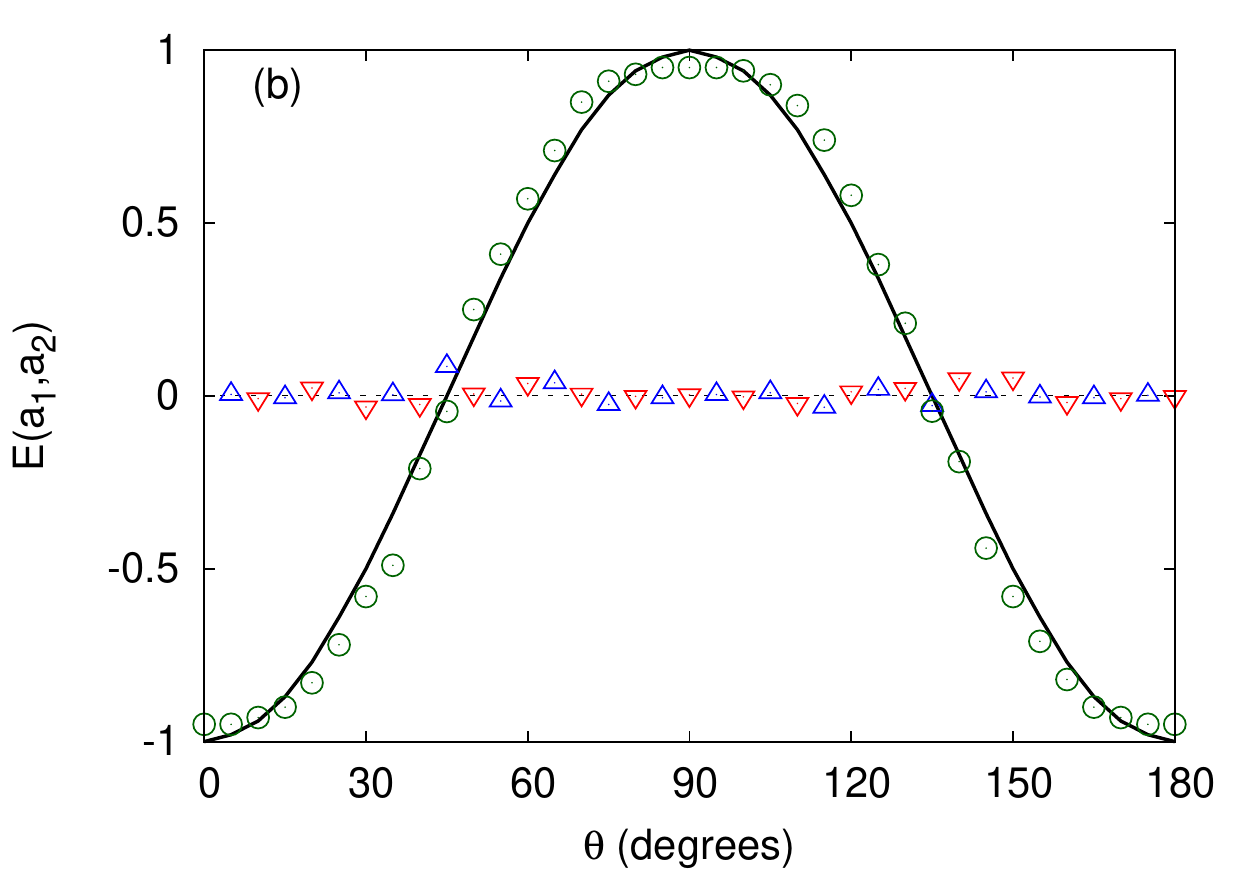}
\caption{%
The correlation $E(a_1,a_2)$ ($\bigcirc$), the single-particle averages $E_1(a_1,a_2)$ ($\bigtriangleup$)
and $E_2(a_1,a_2)$ ($\bigtriangledown$) as a function of $\theta=a_1-a_2$.
(a) CFD-compliant model;
(b) non-CFD model.
Solid line: quantum theoretical prediction $E(a_1,a_2)=-\cos2\theta$.
Dashed line: quantum theoretical prediction $E_1(a_1,a_2)=E_2(a_1,a_2)=0$.
The photon identification threshold is ${\cal V}=-0.995$.
}%
\label{fig4}
\end{center}
\end{figure}

\begin{figure}[t]
\begin{center}
\includegraphics[width=0.45\textwidth]{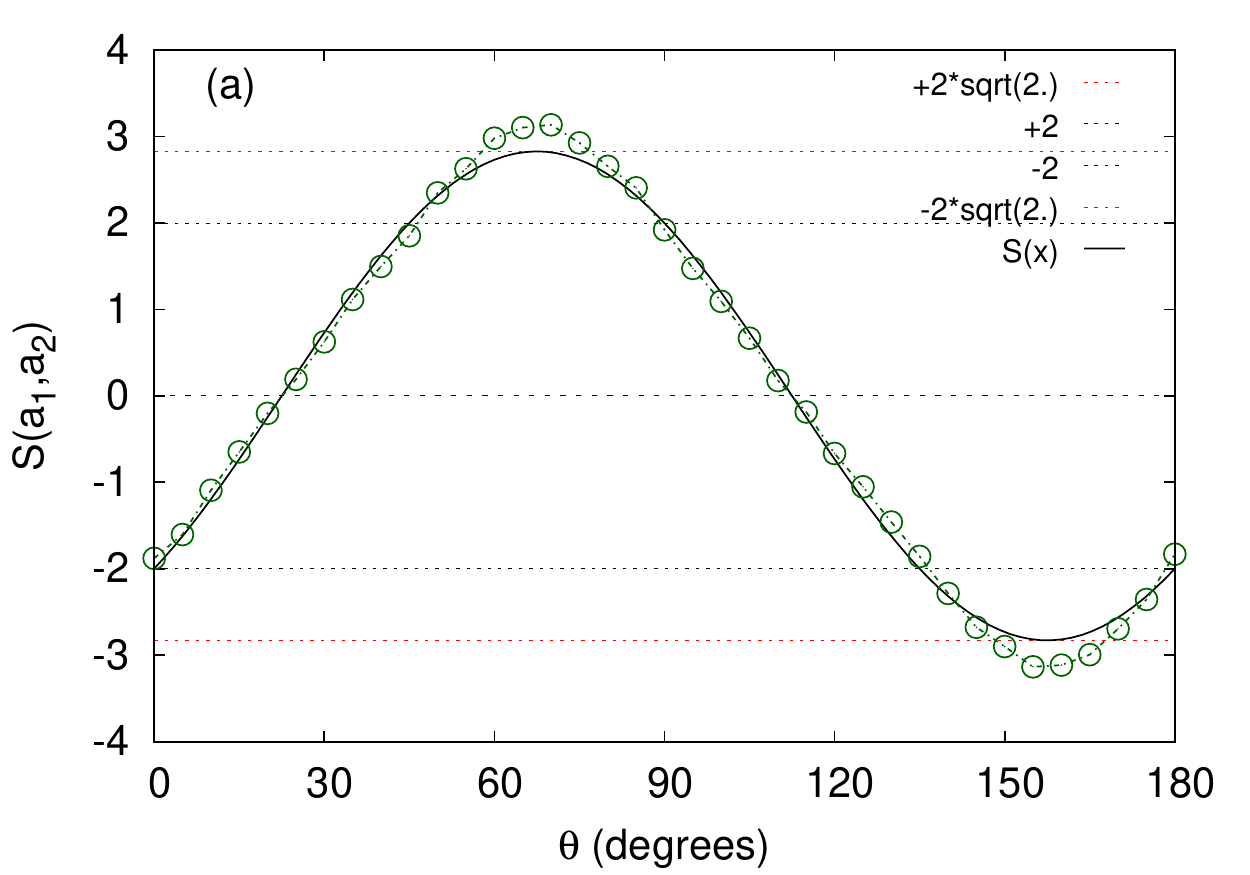}
\includegraphics[width=0.45\textwidth]{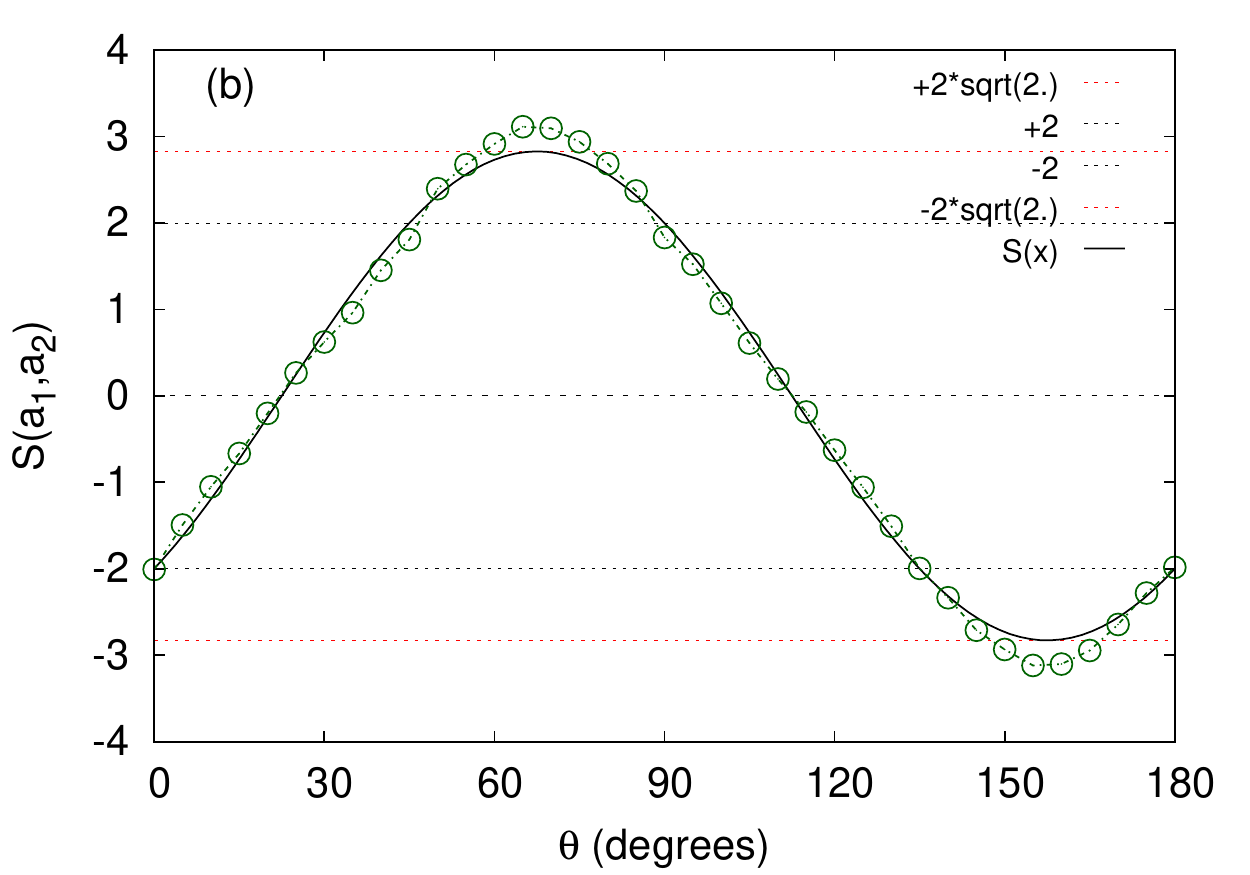}
\caption{%
The Bell-CHSH function $S(\theta+\pi/8,\theta+3\pi/8,\pi/8,3\pi/8)$
obtained from the data shown in Fig.~\ref{fig4} as a function of $\theta=a_1-a_2$.
(a) CFD-compliant model;
(b) non-CFD model.
Solid line: quantum theoretical prediction $S(\theta+\pi/8,\theta+3\pi/8,\pi/8,3\pi/8)=-2\sqrt{2}\cos(2\theta+\pi/4)$.
The photon identification threshold is ${\cal V}=-0.995$.
}%
\label{fig5}
\end{center}
\end{figure}

\begin{figure}[t]
\begin{center}
\includegraphics[width=0.45\textwidth]{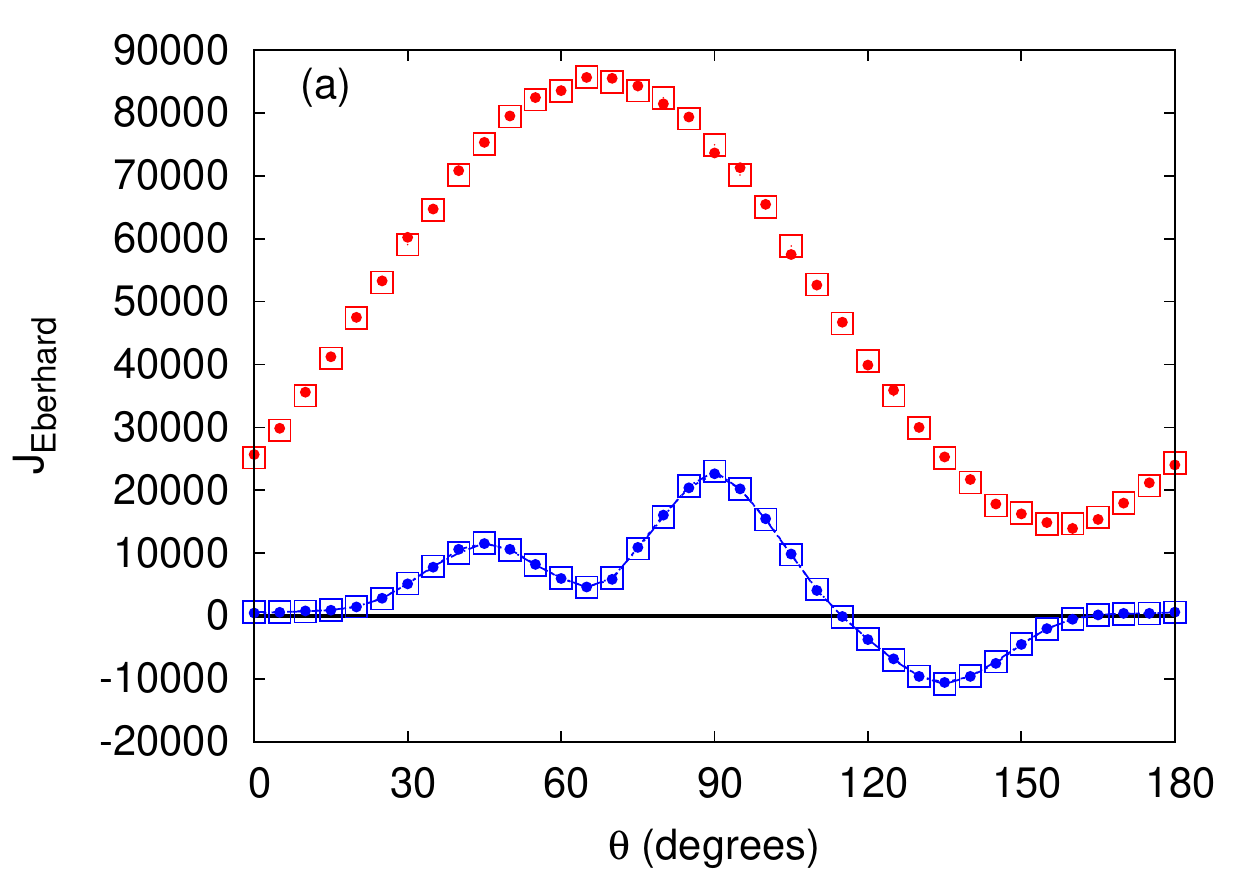}
\includegraphics[width=0.45\textwidth]{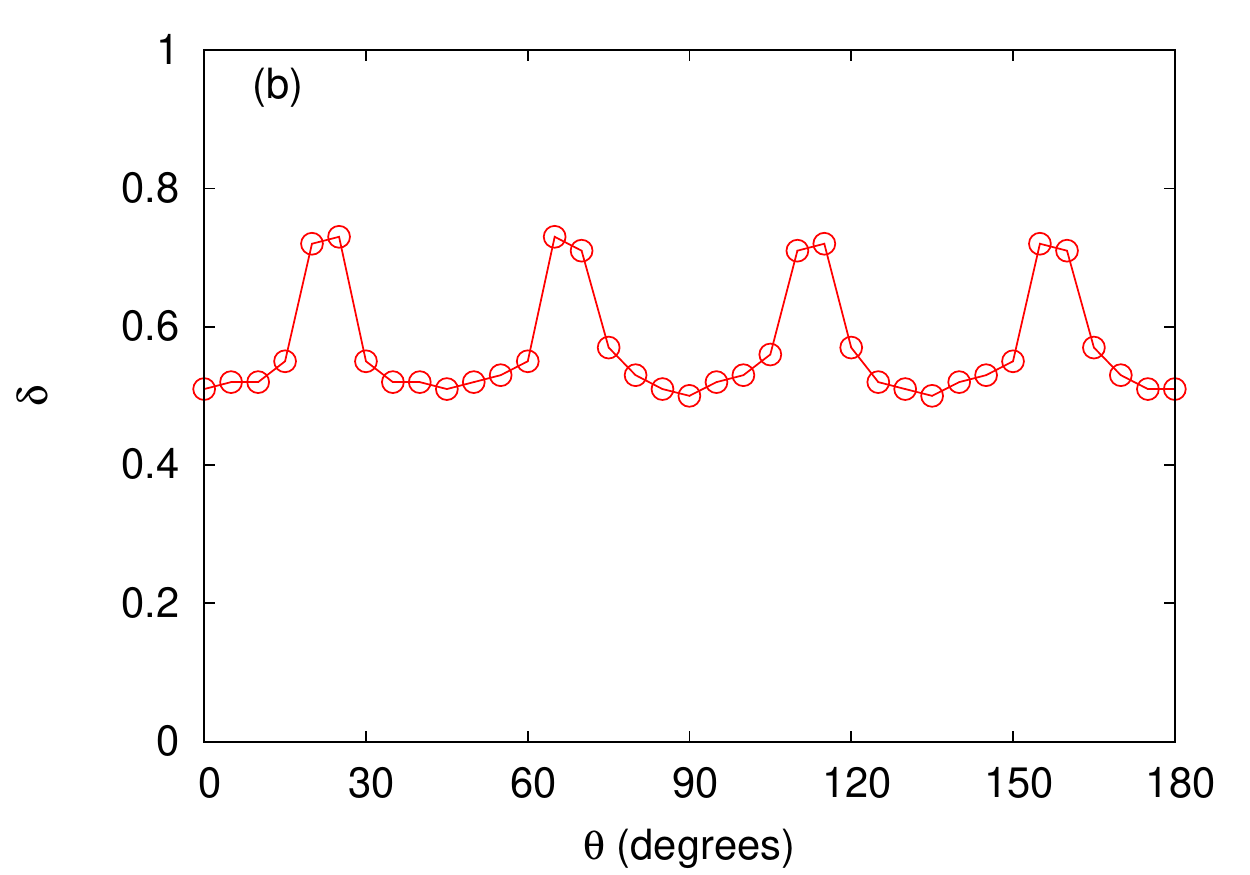}
\caption{%
(a) The Eberhard function $J_\mathrm{Eberhard}$ given by Eq.~(\ref{cfd2b}) as a function of $\theta$,
for the CFD-compliant (open symbols) and non-CFD model (solid symbols), with (circles) and without (squares)
using the photon identification threshold (see text).
(b) The value of $\delta$ entering the inequalities Eq.~(\ref{cfd11}) as a function of $\theta$.
The photon identification threshold ${\cal V}=-0.995$.
}
\label{fig6}
\end{center}
\end{figure}

This section reports the results of simulations with
$N=10^5$ events for the CFD-compliant and $N=10^5$ events per setting for the non-CFD model,
with $V_{\mathrm{min}}=1/2$ and $V_{\mathrm{max}}=1$.
Note that the ``time-tag threshold'' and a ``trigger threshold''
(terminology from Ref.~\cite{GIUS15} (supplementary material)) are important
to the laboratory implementation but are superfluous, meaning that they do not affect the results of our computer experiments in any way.
Indeed, in our perfect experiments, there is no ambiguity in determining when a particle arrives at the observation station.
Nevertheless, to counter possible (pointless) critique that we have not incorporated into our simulation model
the two thresholds that are essential to the laboratory implementation,
we have chosen $V_{\mathrm{min}}=1/2$ in order to leave room for introducing these thresholds.

We limit the discussion to the case $d=4$ because we know from earlier work~\cite{RAED07c,ZHAO08,MICH14a},
which uses time-coincidence selection, that for $d=4$ the computer model reproduces the quantum theoretical result
of the correlation of two particles in the singlet state, Malus' law for the single-particle averages etc.
if $N\rightarrow\infty$ followed by ${\cal V}\rightarrow - V_{\mathrm{max}}$.

It is not difficult to see that single-particle averages $E_1(a_1,a_2)$, $E_2(a_1,a_2)$ etc. are
expected to be zero, up to fluctuations.
The reason is that $\phi \rightarrow \phi+\pi/2$
changes the sign of the $x$'s but has no effect on the values of the $v$'s (see Eq.~(\ref{cfd0b})).
Therefore, if the $\phi's$ uniformly cover $[0,2\pi[$,
the number of times that $x=+1$ and $x=-1$ appear is about the same.
All our simulation results are in concert with this prediction.

In Fig.~\ref{fig4}, we present the simulation data of
the correlation $E(a_1,a_2)$ ($\bigcirc$), the single-particle averages $E_1(a_1,a_2)$ ($\bigtriangleup$)
and $E_2(a_1,a_2)$ ($\bigtriangledown$) as a function of $\theta=a_1-a_2$, as obtained
from a CFD-compliant (Fig.~\ref{fig4}(a)) and non-CFD (Fig.~\ref{fig4}(b)) simulation.
All the simulation data are in excellent agreement with the quantum theoretical description of
a two-particle system in the singlet state which predicts $E_1(a_1,a_2)=E_2(a_1,a_2) =0$ and $E(a_1,a_2) =-\cos2\theta$.
Within statistical fluctuations, it is difficult to distinguish between CFD-compliant and non-CFD
simulation data, in concert with our earlier work~\cite{RAED16c}.

In Fig.~\ref{fig5} we show the data of the Bell-CHSH function $S(\theta+\pi/8,\theta+3\pi/8,\pi/8,3\pi/8)$ as a function of $\theta$.
Clearly the simulation results are in excellent
agreement with the quantum theoretical prediction $S(\theta+\pi/8,\theta+3\pi/8,\pi/8,3\pi/8)=-2\sqrt{2}\cos(2\theta+\pi/4)$.
In both Figs.~\ref{fig4} and Fig.~\ref{fig5}, there are deviations from the quantum theoretical prediction
which are not due to statistical fluctuations.
These deviations can be reduced systematically and eventually vanish
by letting  ${\cal V}\rightarrow -V_{\mathrm{max}}$  (${\cal V}\rightarrow >-V_{\mathrm{max}}$),
a fact that can be proven rigorously for the probabilistic version of the simulation model~\cite{RAED07c,ZHAO08,MICH14a}.

We note in passing that the observation that the frequency distribution of many events agrees with
the probability distribution of a singlet state is a post-factum
characterization of the repeated preparation and measurement process only,
not a demonstration that at the end of the preparation stage, each pair of particles actually {\sl is} in an entangled state.
The latter describes the statistics, not a property of a particular pair of particles~\cite{BALL03}.

Results of the Eberhard function Eq.~(\ref{cfd2b}) as a function of $\theta$ are given in Fig.~\ref{fig6}(a).
The correspondence between the symbols used in Eberhard's
and this paper are as follows: $o\Leftrightarrow+1$, $e\Leftrightarrow-1$,
$\alpha_1\Leftrightarrow a_1^\prime$, $\alpha_2\Leftrightarrow a_1$,
$\beta_1\Leftrightarrow a_2$, and $\beta_2\Leftrightarrow a_2^\prime$.
As expected from the requirements to derive Eq.~(\ref{cfd2b}) (see section~\ref{BEL}),
the CFD-compliant simulation without the photon identification threshold satisfies $J_\mathrm{Eberhard}\ge0$ for all $\theta$
whereas processing the data as in the laboratory experiment, i.e. by employing a photon identification threshold,
yields $J_\mathrm{Eberhard}<0$ for a non-zero interval of $\theta$'s.
As is clear from Fig.~\ref{fig6}(a), the results of Eberhard function Eq.~(\ref{cfd2b}) do
not change significantly if we use replace the CFD-compliant simulation model by its non-CFD version.
The reason for this apparent violation is that the data obtained through the
application of the photon identification mechanism do not satisfy the mathematical requirements for deriving Eq.~(\ref{cfd2b}).

For completeness, in Fig.~\ref{fig6}(b) we present results for the function $\delta$ which determines
the upperbound to the Bell-CHSH function in the case that the photon identification threshold
is being used to discard detection events (see Eq.~(\ref{cfd11})).
From Fig.~\ref{fig6}(b), it follows that $\delta<0.8$. Hence Eq.~(\ref{cfd11}) predicts
an upperbound that is not smaller than $3.4$, large enough to include the maximum
value of $2\sqrt{2}\approx2.83$ predicted by the quantum theory of the polarizations of two photons (or, equivalently, two spin-1/2 particles).

Finally, it is instructive to compare the number of detection events that
the photon identification threshold rejects as being a photon.
In the laboratory experiment~\cite{GIUS15},
the number of trials is about $3.5\times10^9$ and the total number of so-called ``relevant counts'',
i.e. the number of times that at least one photon was identified by means of the photon identification thresholds (by software),
is about $1.8\times10^5$.
Thus, in this experiment the overall number of events considered to be relevant for the
physics, relative to the number of detection events is about $0.005\%$.
For comparison, in the simulations, a photon identification threshold ${\cal  V}=-0.995$ identifies about $23\%$
of the detection events as photons, several orders of magnitude larger than in the laboratory experiment.
Clearly, the quality of the data collected in the laboratory experiments are not on par
with the quality of the data produced by the computer experiments
but obviously, the latter is much easier to realize and use than the former.

\section{Post-factum justification of the simulation model}\label{POS}

\begin{figure}[t]
\begin{center}
\includegraphics[width=0.45\textwidth]{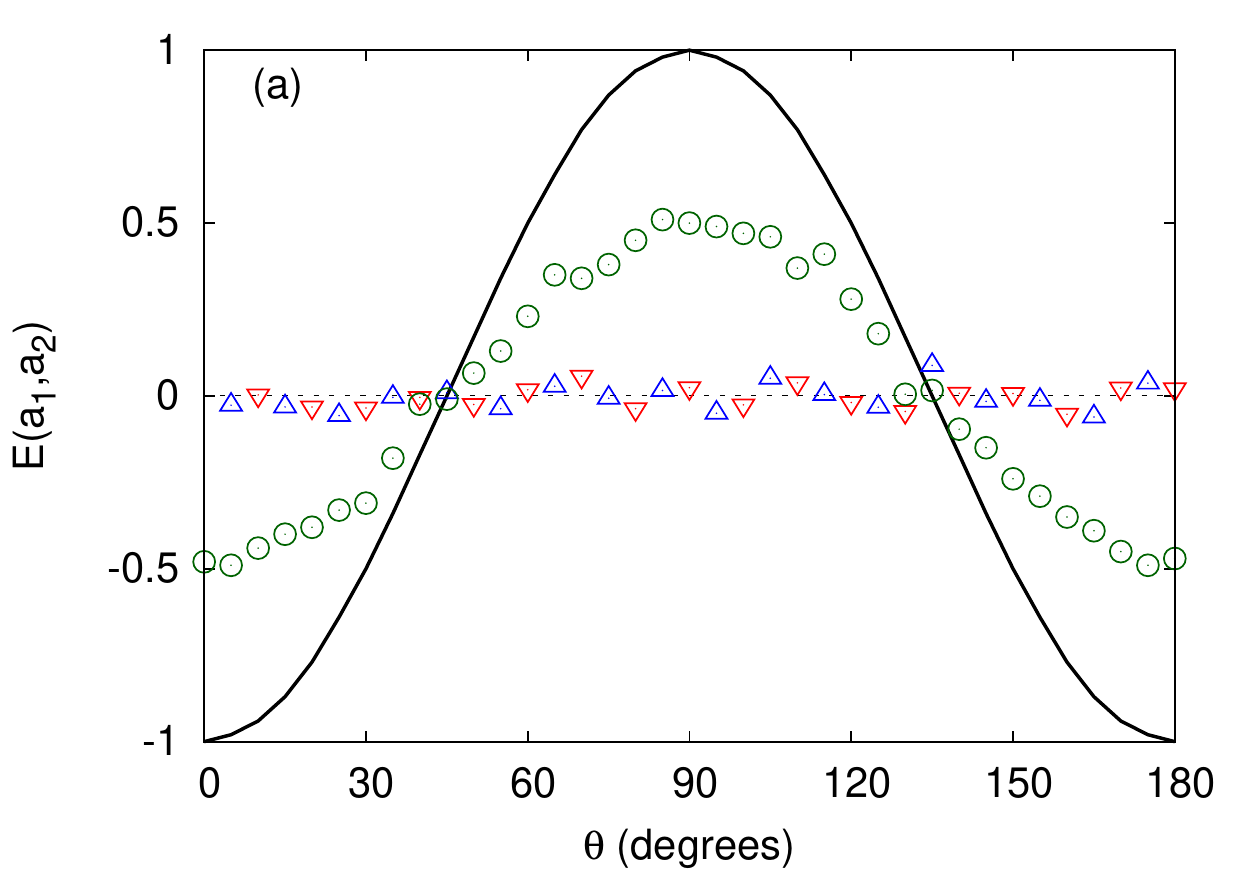}
\includegraphics[width=0.45\textwidth]{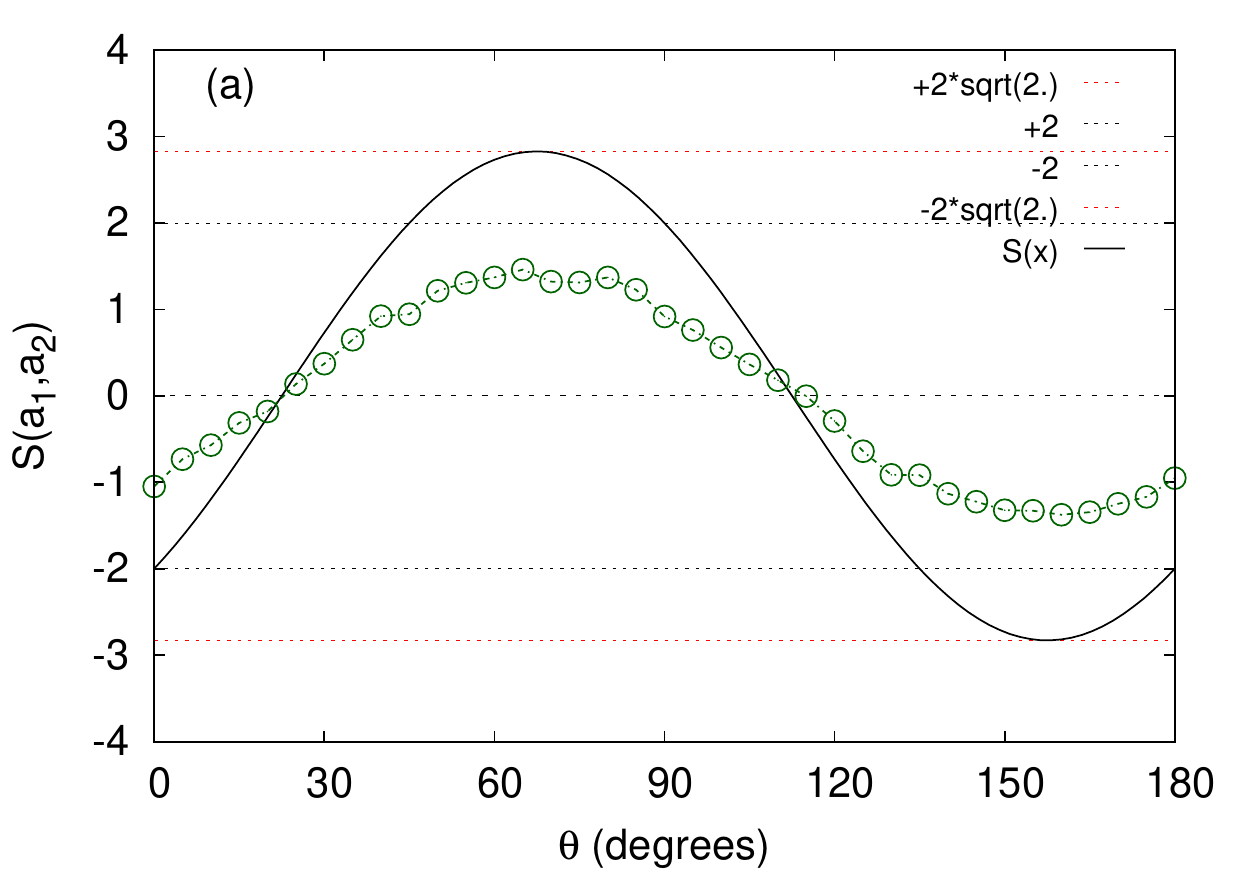}
\caption{%
Demonstration that the simulation model cannot reproduce the correlation of the quantum system
if the voltage signals $v_1$ and $v_2$ are independent of the settings $a_1$ and $a_2$, respectively.
(a) The correlation $E(a_1,a_2)$ ($\bigcirc$), the single-particle averages $E_1(a_1,a_2)$ ($\bigtriangleup$)
and $E_2(a_1,a_2)$ ($\bigtriangledown$) as a function of $\theta=a_1-a_2$ as obtained from a CFD-compliant simulation.
(b) The Bell-CHSH function $S(\theta+\pi/8,\theta+3\pi/8,\pi/8,3\pi/8)$
obtained from the data shown in Fig.~\ref{fig7}(a) as a function of $\theta$.
The photon identification threshold ${\cal V}=-0.995$, $V_\mathrm{min}=0.95$, $V_\mathrm{min}=1$, and $d=0$.
Solid lines: quantum theoretical predictions.
}%
\label{fig7}
\end{center}
\end{figure}

We have already drawn attention to the fact that the explicit expression of the voltage $v=v(a,\phi,\widehat r)$
shows a dependence on the local setting of the station
through the factor $|\sin[2(a-\phi)]|^d$ (see Eqs.~(\ref{cfd0a}) and (\ref{cfd0b}))
and mentioned that such a dependence cannot be ruled out a posteriori.
In this section, we examine the consequences of removing this dependence.

In Fig.~\ref{fig7}, we show results for $d=0$, in which case the random variations of the voltage signals
$v_1$, $v_1^\prime$, $v_2$, and $v_2^\prime$ do not depend on $a_1$, $a_1^\prime$, $a_2$, and $a_2^\prime$, respectively.
Instead of $E(a_1,a_2)\approx -\cos2\theta$ for $d=4$, the simulation for $d=0$ yields $E(a_1,a_2)\approx -(1/2)\cos2\theta$
and, as Fig.~\ref{fig7}(b) shows, $|S(\theta+\pi/8,\theta+3\pi/8,\pi/8,3\pi/8)| \le 2$.
Therefore, the only way to have simulation models of the laboratory experiment~\cite{GIUS15}
reproduce the quantum theoretical prediction of the polarizations of two photons (or, equivalently, two spin-1/2 particles)
is to assume that $v_1$, $v_1^\prime$, $v_2$, and $v_2^\prime$ depend on $a_1$, $a_1^\prime$, $a_2$, and $a_2^\prime$.
Of course, there is no good argument why, in a particular experiment, this dependence should be of the form
Eq.~(\ref{cfd0b}). We repeat that we have chosen the form Eq.~(\ref{cfd0b}) because
our main goal is to reproduce by a CFD-compliant, manifestly non-quantum model,
the quantum theoretical predictions of the polarizations of two photons (or, equivalently, two spin-1/2 particles), which as a by-product,
yields $|S(\theta+\pi/8,\theta+3\pi/8,\pi/8,3\pi/8)|>2$.

Disregarding the original motivation to perform the Bell-test experiments,
the experimental setup shown in Fig.~\ref{fig0} can be regarded as a tool
to characterize the response of the observation stations to the incoming signals.
In the case at hand, what is under scrutiny is the response of the observation station,
i.e. of its optical components, the transition-edge detector and the electronics that amplifies its current,
under the condition that the incident light is extremely feeble.
Viewed from this perspective, our simulations support the hypothesis that the laboratory experiments~\cite{GIUS15,SHAL15}
convincingly demonstrate that the statistics of the observed photons, as defined by the photon detection threshold,
depends on the settings (and hence on the polarizations assigned to the photons) of the observation stations.

It is of interest to mention here that since the early days of the Bell-test experiments,
it is well-known that application of a Bell-type model requires at least one extra assumption.
We reproduce here a passage from Ref.~\cite{CLAU74} (p.1890):
{\it ``The approach used by CHSH is to introduce an auxiliary assumption, that if a particle passes through a spin
analyser, its probability of detection is independent of the analyser's orientation.
Unfortunately, this assumption is not contained in the hypotheses of locality, realism or determinism.''}
It is stunning that although there is at least one auxiliary assumption
involved in testing e.g. the CHSH inequality with Bell-test data, the possibility
that this assumption is not valid is, to the best of our knowledge, ignored in the experimental studies.
As a matter of fact, as we have argued above,  all Bell-test experiments with photons performed up to this day can be regarded as
direct experimental proof that this auxiliary assumption is invalid.
In view of the intricate atomic-scale processes that are involved when light passes through a material,
this conclusion seems very reasonable but is, of course, way less spectacular than the conclusion that
Bell-type experiments can be used to rule out certain world views.

\section{Conclusion}\label{CON}

The general message of this paper is that a model that purports to describe the data produced by an experiment
should account for {\sl all} the data that are relevant for the analysis of the experimental results.
In the case at hand, the situation is as follows:

\begin{enumerate}[(i)]
\item
experimental data~\cite{GIUS15} are interpreted in terms of a Bell-type model that uses only half of the variables (the $x$'s),
\item
in the actual experiments~\cite{GIUS15} the other half of the variables (the $v$'s) is essential for the identification of the photons
but are ignored in Bell-type models,
\item
the failure of Bell-type models to describe the experimental data is taken as a proof that ``local realism'' (local in Einstein's sense)
is incompatible with quantum theory and is therefore is declared dead.
\end{enumerate}
We believe that it requires an exotic form of logic to reconciliate the last statement (iii) with the second one (ii).

To head off possible misunderstandings,
the authors of this paper do not necessarily subscribe to all or any forms of what is called local realism, CFD theories, or ...
We are of the opinion that the arguments based on Bell's theorem in conjunction with Bell-type experiments
suffer from what we earlier called the photon identification loophole.
One simply cannot blame a model that only accounts for part of the data for not describing all of them.
Regarding the previous sentence,
Albert Einstein's quote ``make it as simple as possible, but not simpler'' is more pertinent than ever.

The challenge for the Bell-experiments community is, therefore, to construct an
EPRB-type experiment with a photon (pair) identification that cannot, from the perspective of the simple
Bell models, be turned into a ``loophole''.
Our general proofs of the derivation of Bell-type inequalities for actual data (see section~\ref{BEL}),
indicate that this challenge cannot be met.

\section*{Acknowledgements}
We like to thank D. Willsch, F. Jin, and M. Nocon for useful comments and discussions.

\appendix*
\section{Probabilistic models}\label{EAR}

Traditionally, mathematical models of the EPRB experiments are formulated in terms of
probabilistic models~\cite{CLAU69,PEAR70,FINE74,BRAN87,PENA72,FINE82,FINE82a,FINE82b,MUYN86,KUPC86,JAYN89,%
BELL93,BROD93,PITO94,FINE96,KHRE99,SICA99,BAER99,HESS01a,HESS01b,HESS05,ACCA05,KRAC05,SANT05,LOUB05,%
KUPC05,MORG06,KHRE07,ADEN07,NIEU09,MATZ09,KARL09,KHRE09,GRAF09,KHRE11,NIEU11,RAED11a,HESS15,KOFL16},
often without explicitly mentioning Kolmogorov's axiomatic framework of probability theory~\cite{KOLM56,GRIM95}.
However, there is a considerable,  conceptual gap between a laboratory EPRB experiment and a probabilistic model thereof.
The (over)simplifications required to come up with a tractable, proper probabilistic model of a laboratory EPRB experiment
are key to the understanding of the phenomena involved.

When we use the ``digital computer -- laboratory experiment'' metaphor,
both the simplification and replacement are made during the formulation of the computer model.
A computer simulation algorithm entails a complete specification of how the data are generated.
In this respect, all ``physically relevant'' processes are well-defined
(by construction) and known explicitly in full detail.
There are no uncertainties or unknown influences.
Note that the reverse operation, i.e. to construct an algorithm for a digital computer
out of a probabilistic model is, as a matter of principle, impossible.
At most, a probabilistic model can serve as a guide to construct an algorithm.

In the particular case where the observed phenomena take the form of
data generated by simulations of EPRB experiments on a digital computer,
the transition from the observed phenomena to suitable mathematical models
does not suffer from the many ``uncertain'' factors that may or may
not play an essential role in the laboratory experiments
and is, therefore, a fairly simple transition.
In this section, we start from a computer simulation model and construct the probabilistic model thereof.
We start by showing that the simplest of these models are identical to those proposed and analyzed by Bell~\cite{BELL01}
and then move on to the construction of a probabilistic model for the computer model
of the recent Bell-type experiments~\cite{GIUS15,SHAL15} that we use in our simulation work presented in section~\ref{RES}.

\subsection{Bell-type models}\label{SIM}

Consider the CFD simulation model in which we explicitly ignore the $v$-variables.
For a given input to the observation stations, the outcome is one of the so-called elementary events~\cite{KOLM56,GRIM95},
in this case one of the $16$ different quadruples $(x_{1},x_{1}^\prime,x_{2},x_{2}^\prime)$.
In the language of probability theory, the set of these 16 different quadruples is called the sample space $\Omega$~\cite{KOLM56,GRIM95},
the set of elementary events, from which we construct the so-called $\sigma$-field ${\cal F}$ of subsets of $\Omega$,
containing the impossible (null) event and all the (compound) events in whose occurences we may be interested~\cite{KOLM56,GRIM95}.
In modeling the computer experiments, we only need to consider finite sets, hence we do not have to worry about
the mathematical subtleties that arise when dealing with infinite sets~\cite{KOLM56,GRIM95}.

The next step is to assign a real number, a probability measure, between zero and one that expresses
the likelihood that an element of the set $\Omega$ occurs~\cite{KOLM56,GRIM95}.
We denote this (conditional) probability measure by $P(x_{1},x_{1}^\prime,x_{2},x_{2}^\prime|a_1,a_1^\prime,a_2,a_2^\prime, Z)$,
the part $|a_1,a_1^\prime,a_2,a_2^\prime, Z)$ indicating that
the settings and all other conditions, denoted collectively by $Z$, do not change during the imaginary probabilistic experiment.
By definition, the probability measure satisfies
$\sum_{(x_{1},x_{1}^\prime,x_{2},x_{2}^\prime)\in\Omega} P(x_{1},x_{1}^\prime,x_{2},x_{2}^\prime|a_1,a_1^\prime,a_2,a_2^\prime, Z) =1$~\cite{KOLM56,GRIM95}.

Note that a probability measure is a purely mental construct~\cite{KOLM56}.
If it were not, we could interchange the experiment/computer simulation, the results of which directly connect to our senses,
with the imaginary world of mathematical models and prove theorems,
not only about the mathematical description, but also about our sensory experiences, a tantalizing possibility.
One such example that exploits intricate features of set theory is given in Ref.~\cite{PITO82},
in which it is explicitly stated that there does not exist an algorithm to actually calculate
the relevant functions.
In other words, this example cannot be realized on a physical device such as a digital computer, not even approximately.
Moreover, unlike the simulation algorithm executing on a digital computer,
the probabilistic description does not contain a specification of the process that actually produces an event:
we have to call up Tyche to do this for us.
In other words, a probabilistic model is incomplete in that it only describes the outcomes of the simulation procedure,
not the procedure itself. However, this incompleteness is partially compensated for by the fact that the calculation of averages
and correlations no longer involves the number of events $N$.
We have for instance
\begin{eqnarray}
\widetilde E_{12}(a_1,a_1^\prime,a_2,a_2^\prime)&=&\sum_{\Omega}
x_{1}x_{2}P(x_{1},x_{1}^\prime,x_{2},x_{2}^\prime|a_1,a_1^\prime,a_2,a_2^\prime, Z)
,
\label{ear3a}
\end{eqnarray}
and we have similar expressions for the other two-particle averages and also for the single-particle averages.
Here and in the following, we use the shorthand notation $\sum_{\Omega}=\sum_{(x_{1},x_{1}^\prime,x_{2},x_{2}^\prime)\in\Omega}$.
We have written $\widetilde{E}$ instead of $\widehat{E}$ to emphasize that
the former have been calculated within a probabilistic model whereas the latter involve calculations with actual data.
From Eq.~(\ref{ear3a}), we have
\begin{eqnarray}
\widetilde S(a_1,a_2,a_1^\prime,a_2^\prime)&=&\sum_{\Omega}
\left( x_{1}x_{2} - x_{1}^\prime x_{2} + x_{1}x_{2}^\prime + x_{1}^\prime x_{2}^\prime \right)
\nonumber \\
&&\hbox to 1cm{} \times P(x_{1},x_{1}^\prime,x_{2},x_{2}^\prime|a_1,a_1^\prime,a_2,a_2^\prime, Z)
,
\label{ear3b}
\end{eqnarray}
and because the elementary events are quadruples, it follows directly from Eq.~(\ref{cfd1a}) that
$|\widetilde S(a_1,a_2,a_1^\prime,a_2^\prime)|\le 2$.
Thus, {\it in the probabilistic realm}, not in the world of the observed two-valued data,
the existence of the Bell-CHSH inequality follows from the existence of a probability measure
for the elementary events of quadruples $(x_{1},x_{1}^\prime,x_{2},x_{2}^\prime)$.
Moreover, it can be shown that with some additional requirements on its marginals, the existence of such a probability measure
is necessary and sufficient for Bell-type inequalities to hold~\cite{VORO62,SUPP81,FINE82a,BROD89,PITO91}.

It is clear from Fig.~\ref{fig3} that $x_i$ and $x_i^\prime$ for $i=1,2$ only depend on the corresponding
$a_i$ and $a_i^\prime$, respectively.
However, the probability measure for quadruples, $P(x_{1},x_{1}^\prime,x_{2},x_{2}^\prime|a_1,a_1^\prime,a_2,a_2^\prime, Z)$,
does not express this basic property of the computer model,
nor does it explicitly express the dependence on the $\phi$'s.

A simple way to incorporate all these features of the simulation model in a
probabilistic description is to define a new joint probability measure for quadruples by
\begin{eqnarray}
\lefteqn{
P^\prime(x_{1},x_{1}^\prime,x_{2},x_{2}^\prime|a_1,a_1^\prime,a_2,a_2^\prime, Z) =}
\nonumber \\
&&\int
P(x_{1}|a_1,\hidden_1, Z)
P(x_{1}^\prime|a_1^\prime,\hidden_1, Z)
P(x_{2}|a_2,\hidden_2, Z)
\nonumber \\
&&\hbox to 2cm{}\times
P(x_{2}^\prime|a_2^\prime,\hidden_2, Z)
\mu(\hidden_1,\hidden_2) d\hidden_1 d\hidden_2
,
\label{ear3c}
\end{eqnarray}
where $P(x_{1}|a_1,\hidden_1, Z)$ etc. are the ``local'' probabilities
to observe $x_{1}$ etc.,
the integration is over
the whole domain of $\hidden_1$ and $\hidden_2$ and $\mu(\hidden_1,\hidden_2)$
is a non-negative, normalized density.
With the new probability measure Eq.~(\ref{ear3c}), Eq.~(\ref{ear3a}) simplifies considerably.
For instance, for the {\bf detection events} we have
$\widetilde E_{12}(a_1,a_1^\prime,a_2,a_2^\prime)=\widetilde E(a_1,a_2)$ where
\begin{eqnarray}
\widetilde E(a,b)
=\sum_{x,y=\pm1} \int xy P(x|a,\hidden_1, Z)P(y|b,\hidden_2, Z) \mu(\hidden_1,\hidden_2) d\hidden_1 d\hidden_2
.
\label{ear3e}
\end{eqnarray}
Instead of Eq.~(\ref{ear3b}), we now have
\begin{eqnarray}
|\widetilde S(a_1,a_2,a_1^\prime,a_2^\prime)|=
|\widetilde E(a_1,a_2)-\widetilde E(a_1,a_2^\prime)+\widetilde E(a_1^\prime,a_2)+\widetilde E(a_1^\prime,a_2^\prime)|\le 2
,
\label{ear3f}
\end{eqnarray}
which is the Bell-CHSH inequality in probabilistic form~\cite{CLAU69,BELL93}.

From Eq.~(\ref{ear3c}), it follows directly that
\begin{eqnarray}
P^\prime(x_{1},x_{2}|a_1,a_2, Z)
&=&
\sum_{x_{1}^\prime=\pm1}
\sum_{x_{2}^\prime=\pm1}
P^\prime(x_{1},x_{1}^\prime,x_{2},x_{2}^\prime|a_1,a_1^\prime,a_2,a_2^\prime, Z)
\nonumber \\
&=&
\int
P(x_{1}|a_1,\hidden_1, Z)
P(x_{2}|a_2,\hidden_2, Z)
\mu(\hidden_1,\hidden_2) d\hidden_1 d\hidden_2
,
\nonumber \\
\label{ear3g}
\label{ear3h}
\end{eqnarray}
which expresses the probability measure $P^\prime(x_{1},x_{2}|a_1,a_2, Z)$
in terms of the single-variable probability measures
$P(x_{1}|a_1,\hidden_1, Z)$ and $P(x_{2}|a_2,\hidden_2, Z)$ and the measure
$\mu(\hidden_1,\hidden_2)$ of the variables $\hidden_1$ and $\hidden_2$.

The factorized form Eq.~(\ref{ear3h}) is the landmark of the so-called ``local hidden-variable models'' introduced by Bell~\cite{BELL93}.
Although ``local'' is often used to express the notion that physical influences do not travel faster than the speed of light
it is, in the context of a probabilistic model (computer model), an expression of statistical (arithmetic) independence only.
Bell's theorem uses the factorized form Eq.~(\ref{ear3h}) to state that quantum mechanics is incompatible with local realism,
the world view in which physical properties of objects exist independently of
measurement and where physical influences cannot travel faster than the speed of light~\cite{GIUS15}.

In one respect, Eq.~(\ref{ear3h}) is grossly deceiving, namely
it does not reflect the elementary fact that the parent probability measure Eq.~(\ref{ear3c})
from which Eq.~(\ref{ear3h}) follows, concerns quadruples, not pairs.
Without the knowledge that Eq.~(\ref{ear3h}) is in fact a marginal distribution of the
probability measure Eq.~(\ref{ear3c}) for quadruples, one is inclined to think, as Bell did and his followers still seem to do,
that there are ``physical'' assumptions involved in justifying the factorized form Eq.~(\ref{ear3h}).
However, this is not the case because the Bell-type inequalities hold
{\bf if and only if} there exists a joint probability measure for the quadruples~\cite{FINE82a}.
This mathematical statement is void of any physical meaning.

In summary: in this subsection we have shown that a probabilistic model of the CFD computer simulation model
that does not account for the photon identification mechanism of the EPRB laboratory experiment,
automatically leads to the models introduced by Bell~\cite{BELL93}.
Within this framework, the existence of Bell-type inequalities and the corresponding joint probability measures
are mathematically equivalent~\cite{FINE82a}.
The latter statement, which relates to imaginary data only, corresponds to the statement that
in the realm of actual two-valued data, the existence of Bell-type inequalities and CFD-compliant generation
of all the quadruples are mathematically equivalent, see section~\ref{BEL}.

\subsection{Incorporating the photon identification threshold}\label{INC}

Referring to Eq.~(\ref{fig3}), the extension of the construction outlined in section~\ref{SIM}
to incorporate the local photon identification mechanism is of purely technical nature.
Instead of Eq.~(\ref{ear3c}), we now introduce
\begin{eqnarray}
\lefteqn{
P''(x_{1},v_{1},x_{1}^\prime,v_{1}^\prime,x_{2},v_{2},x_{2}^\prime,v_{2}^\prime|a_1,a_1^\prime,a_2,a_2^\prime, Z) =}
\nonumber \\
&&\int
P(v_{1}|a_1,\hidden_1, Z)
P(x_{1}|a_1,\hidden_1, Z)
P(v_{1}^\prime|a_1^\prime,\hidden_1, Z)
P(x_{1}^\prime|a_1^\prime,\hidden_1, Z)
\nonumber \\&&\hbox to 1.5cm{}\times
P(v_{2}|a_2,\hidden_2, Z)
P(x_{2}|a_2,\hidden_2, Z)
P(v_{2}^\prime|a_2^\prime,\hidden_2, Z)
P(x_{2}^\prime|a_2^\prime,\hidden_2, Z)
\nonumber \\&&\hbox to 2cm{}\times
\mu(\hidden_1,\hidden_2) \;d\hidden_1\;d\hidden_2\;dv_{1}\;dv_{2}
,
\label{inc0}
\end{eqnarray}
where $P(v_{1}|a_1,\hidden_1, Z)$ etc. are the ``local'' probability densities to pick $v_{1}$ etc.,
and all other symbols have the same meaning as in Eq.~(\ref{ear3c}).

It is now straightforward to write down the probabilistic expressions that incorporate in exactly
the same manner as in the analysis of the laboratory experiment data~\cite{GIUS15},
the effect of the photon identification threshold.
For instance, we have for the {\bf photon} counts
\begin{eqnarray}
\widetilde E(a_1,a_2)&=&\frac{A(a_1,a_2)}{B(a_1,a_2)}
,
\nonumber \\
\noalign{\noindent where}
A(a_1,a_2)&=&
\sum_{x_{1},x_{2}=\pm1}\int
x_1x_2
\Theta({\cal V}-v_{1})
\Theta({\cal V}-v_{2})
P(v_{1}|a_1,\hidden_1, Z)
\nonumber \\&&\hbox to 1.5cm{}\times
P(x_{1}|a_1,\hidden_1, Z)
P(v_{2}|a_2,\hidden_2, Z)
P(x_{2}|a_2,\hidden_2, Z)
\nonumber \\&&\hbox to 2cm{}\times
\mu(\hidden_1,\hidden_2) \;d\hidden_1\;d\hidden_2\;dv_{1}\;dv_{2}
,
\nonumber \\
\noalign{\noindent and}
B(a_1,a_2)&=&
\int
\Theta({\cal V}-v_{1})
\Theta({\cal V}-v_{2})
P(v_{1}|a_1,\hidden_1, Z)
P(v_{2}|a_2,\hidden_2, Z)
\nonumber \\&&\hbox to 2cm{}\times
\mu(\hidden_1,\hidden_2) \;d\hidden_1\;d\hidden_2\;dv_{1}\;dv_{2}
,
\label{inc1}
\end{eqnarray}
and, as before, we have similar expressions for the other expectations in Eq.~(\ref{ear3a}) and for the single-particle averages.

The expressions for the single- and two-particles averages that derive from the
probabilistic model Eq.~(\ref{inc0}) all have the form that is
characteristic of a genuine ``local hidden-variable model'', as exemplified by Eq.~(\ref{inc1}).
Only ``local'' detection and photon identification are involved.
The values of the variables local to one observation station do not depend on variables that are local to another observation station.
The only form of ``communication'' between the stations is through the ``hidden'' variables $\hidden_1$ and $\hidden_2$.

It directly follows from the general discussion of section~\ref{EQU} that $A(a_1,a_2)$ and $B(a_1,a_2)$
can be expressed in terms of both the local time-window  and time-coincidence selection.
In detail, for the local time-window selection we have
\begin{eqnarray}
A(a_1,a_2)&=&
\sum_{x_{1},x_{2}=\pm1}
\int
x_1x_2
\Theta(W-t_{1})
\Theta(W-t_{2})
\Theta(t_{1})
\Theta(t_{2})
P(t_{1}|a_1,\hidden_1, Z)
\nonumber \\&&\hbox to 1.5cm{}\times
P(x_{1}|a_1,\hidden_1, Z)
P(t_{2}|a_2,\hidden_2, Z)
P(x_{2}|a_2,\hidden_2, Z)
\nonumber \\&&\hbox to 2cm{}\times
\mu(\hidden_1,\hidden_2) \;d\hidden_1\;d\hidden_2\;dt_{1}\;dt_{2}
,
\nonumber \\
\noalign{\noindent and}
B(a_1,a_2)&=&
\int
\Theta(W-t_{1})
\Theta(W-t_{2})
\Theta(t_{1})
\Theta(t_{2})
P(t_{1}|a_1,\hidden_1, Z)
P(t_{2}|a_2,\hidden_2, Z)
\nonumber \\&&\hbox to 2cm{}\times
\mu(\hidden_1,\hidden_2) \;d\hidden_1\;d\hidden_2\;dt_{1}\;dt_{2}
,
\label{inc2}
\end{eqnarray}
for the time-coincidence selection we have
\begin{eqnarray}
A(a_1,a_2)&=&
\sum_{x_{1},x_{2}=\pm1}
\int
x_1x_2
\Theta(W-|t_{1}-t_{2}|)
P(t_{1}|a_1,\hidden_1, Z)
\nonumber \\&&\hbox to 1.5cm{}\times
P(x_{1}|a_1,\hidden_1, Z)
P(t_{2}|a_2,\hidden_2, Z)
P(x_{2}|a_2,\hidden_2, Z)
\nonumber \\&&\hbox to 2cm{}\times
\mu(\hidden_1,\hidden_2) \;d\hidden_1\;d\hidden_2\;dt_{1}\;dt_{2}
,
\nonumber \\
\noalign{\noindent and}
B(a_1,a_2)&=&
\int
\Theta(W-|t_{1}-t_{2}|)
P(t_{1}|a_1,\hidden_1, Z)
P(t_{2}|a_2,\hidden_2, Z)
\nonumber \\&&\hbox to 2cm{}\times
\mu(\hidden_1,\hidden_2) \;d\hidden_1\;d\hidden_2\;dt_{1}\;dt_{2}
.
\label{inc3}
\end{eqnarray}

From earlier work based on representation Eq.~(\ref{inc3})~\cite{RAED06c,RAED07c,MICH14a}
and from the simulation results presented in section~\ref{RES},
it follows directly that the probabilistic model defined by Eq.~(\ref{inc0}) is
capable of reproducing the predictions of quantum theory
for the single- and two-particles averages of two photon polarizations in the singlet state.
This then should stop spreading the misconception that Bell has proven
that quantum theory is incompatible with {\sl all} ``local hidden-variable models''.
Of course, ``local hidden-variable models'' that do not include the, for the laboratory
experiment essential, mechanism to identify single or pairs of photons are incapable
to describe the salient features of the experimental data but
as explained in section~\ref{MOD}, that is hardly more than a platitude.

In summary, in this appendix we have shown how to construct
probabilistic descriptions of computer simulation models of EPRB experiments
that rely on photon identification thresholds to decide whether or not a photon has been detected~\cite{GIUS15}.
The resulting probabilistic models conform to the requirements of genuine ``local hidden-variable models''
and if they account for a local mechanism to identify photons,
are also capable of producing results that are in full agreement with quantum theory.

\end{document}